\documentclass[aps,pre,reprint,superscriptaddress]{revtex4-2}

\usepackage{amsmath,amsfonts,amssymb,bm,hyperref,graphicx}
\usepackage{longtable}
\usepackage{physics}
\newcommand{\ud}{\mathrm{d}}
\newcommand{\ue}{\mathrm{e}}
\newcommand{\ui}{\mathrm{i}}
\begin{document}

\title{Exploring quasiprobability approach to quantum work in the presence of initial coherence: Advantages of the Margenau-Hill distribution}
\author{Ji-Hui Pei}
\affiliation{School of Physics, Peking University, Beijing, 100871, China}
\author{Jin-Fu Chen}
\thanks{Corresponding author: chenjinfu@pku.edu.cn}
\affiliation{School of Physics, Peking University, Beijing, 100871, China}
\author{H. T. Quan}
\thanks{Corresponding author: htquan@pku.edu.cn}
\affiliation{School of Physics, Peking University, Beijing, 100871, China}
\affiliation{Collaborative Innovation Center of Quantum Matter, Beijing 100871,
China}
\affiliation{Frontiers Science Center for Nano-optoelectronics, Peking University,
Beijing, 100871, China}
\date{\today}

\begin{abstract}
    In quantum thermodynamics, the two-projective-measurement (TPM) scheme provides a successful description of stochastic work only in the absence of initial quantum coherence. 
    Extending the quantum work distribution to quasiprobability is a general approach to characterize work fluctuation in the presence of initial coherence. 
    However, among a large number of different definitions, there is no consensus on the most appropriate work quasiprobability. 
    In this article, we list several physically reasonable requirements including the first law of thermodynamics, time-reversal symmetry, 
    positivity of second-order moment, and a support condition for the work distribution. 
    We prove that the only definition that satisfies all these requirements is the Margenau-Hill (MH) quasiprobability of work. 
    In this sense, the MH quasiprobability of work shows its advantages over other definitions. 
    As an illustration, we calculate the MH work distribution of a breathing harmonic oscillator with initial squeezed states and 
    show the convergence to classical work distribution in the classical limit. 
\end{abstract}
\maketitle

\section{Introduction}
In stochastic thermodynamics, the studies of thermodynamics were extended from the ensemble average level to the stochastic trajectory level \cite{Seifert_2012,book_ST}, 
where work and other thermodynamic concepts are identified as stochastic quantities \cite{sekimoto2010stochastic}. 
These stochastic quantities were found to satisfy some interesting and strong relations known as fluctuation theorems \cite{FTs}. 
On even more microscopic scales, quantum effects play a crucial role in thermodynamic phenomena \cite{Sai2016,Pekola2015,book_ox,Goold_2016,PhysRevX.8.011019,Campisi_2015,Millen_2016,PhysRevX.8.031037}. 
Due to the absence of the concept of trajectory in quantum mechanics, it is a necessary but challenging task to define the quantum stochastic work \cite{PhysRevE.93.022131}. 
After some early efforts on the operator-of-work scheme \cite{o-of-work_77,o-of-work_05}, 
the stochastic work defined by the two-projective-measurement (TPM) scheme \cite{TPM1,TPM2} is now the standard way provided no quantum coherence in the initial state. 
Based on this definition, the fluctuation theorems are readily recovered when the quantum system is driven from canonical equilibrium. 
The TPM scheme is also widely used to define other quantum thermodynamic quantities, including heat and particle number exchange \cite{Talkner_2009}. 
The consequent quantum thermodynamic theory has flourished in both theoretical and experimental aspects 
\cite{Sai2016,Pekola2015,book_ox,Goold_2016,PhysRevX.8.011019,Campisi_2015,Millen_2016,PhysRevX.8.031037,RevModPhys.81.1665,RevModPhys.83.771}. 

However, the TPM scheme fails to characterize the quantum stochastic work when the initial state of the system contains quantum coherence, 
that is, the initial density matrix does not commute with the initial Hamiltonian. 
The first measurement process becomes invasive and destroys the coherence of the initial state \cite{exp_work,An2015}. 
Quantum coherence indeed brings some advantages in quantum thermodynamics. 
It has been shown that quantum coherence can improve the performance of quantum heat engines \cite{QHE_Carnot,QHE_linear,QHE_experiment,QHE_extract,many_cycles,MYH,otto_1,otto_2}, 
and can be used to enhance the work extraction \cite{extract_correlation,extract_coherence}. 
To characterize the fluctuation with initial coherence, a generalized definition of stochastic work is needed, 
which should recover the standard result based on the TPM scheme if there is no initial coherence. 
Unfortunately, the no-go theorem for quantum work prohibits a genuine probability framework to reconcile the first law and the reproduction of the TPM result \cite{nogo}. 
A possible resolution is to adopt the quasiprobability for the work distribution, which allows the distribution to take negative or complex values. 
It is found that the work quasiprobability is a direct signature of quantum contextuality \cite{Contextuality}. 

Quasiprobability distributions arise naturally in the study of quantum mechanics when treating noncommutative observables \cite{Wigner,o_K,o_D,o_MH}. 
Recently, there is growing interest in studying quasiprobability of work in quantum thermodynamic phenomena \cite{QW_Ising,KDQ_fermionic}. 
So far, there have been several different proposals on the definitions of quantum work. 
For example, the Kirkwood-Dirac (KD) quasiprobability of quantum work \cite{w_KD,w_SE,w_SE2} and the Margenau-Hill (MH) quasiprobability of work \cite{w_MH} 
are based on famous quasiprobability distributions \cite{o_MH,o_K}, which are widely used in quantum optics. 
The Full-Counting (FC) quasiprobability of work \cite{w_FCS} is another kind of quasiprobability, which is related to full counting statistics \cite{o_FCS}. 
Most recently, a class of quasiprobability distributions that generalizes the MH and FC ones was proposed \cite{family}. 
A natural question arises: which definition of quantum work quasiprobability is the most appropriate one?

In this article, we propose several requirements for quantum work 
and further show that only the MH quasiprobability of work meets all these requirements. 
Hence, it is the most appropriate definition of quantum stochastic work. 
This paper is organized as follows. In Sec. \ref{review}, we first briefly review the TPM scheme and the no-go theorem for quantum work. 
In Sec. \ref{s_requirements}, we list the requirements for an appropriate quantum work and discuss their physical implications. 
Among these requirements, the support condition is the most conspicuous one. 
In Sec. \ref{proof}, we prove the only definition that satisfies all the requirements is the MH quasiprobability of work. 
In Sec. \ref{discussion}, a comparison of some definitions is discussed from the perspective of these requirements. 
In Sec. \ref{s_example}, we calculate the MH work quasiprobability for a breathing quantum harmonic oscillator starting from a squeezed state 
and show the quantum-classical correspondence principle for MH work distribution.

\section{A review of no-go theorem for quantum work}\label{review}
The isolated driving process is one of the basic processes in the study of nonequilibrium statistical mechanics. 
Let us consider a nonequilibrium quantum driving process governed by the Schrödinger equation during the time interval $[0,\tau]$.  
The time-dependent Hamiltonian is $H_S(t),0\leq t\leq \tau$, 
and the initial density matrix for the system is denoted as $\rho$. 
Here we use the subscript $S$ in the Hamiltonian to denote the Schrödinger picture. 
For simplicity, we set the reduced Planck constant $\hbar=1$ except in Sec. \ref{s_example}. 
The time evolution of the system is characterized by the time-evolution operator $U(\tau)=\mathcal T \exp(-\ui\int_0^\tau\ud t H_S(t))$, where $\mathcal T$ is the time-ordering operator. 
In the Heisenberg picture, the Hamiltonian operator is $H_H(t)= U(t)^{-1} H_S(t) U(t)$ while the quantum state remains invariant. 
In the later discussion, we will mostly work in the Heisenberg picture 
and adopt several shorthand notations: $ H_0= H_H(0)$, $H_1=H_H(\tau)$, $U=U(\tau)$. 
We diagonalize the initial and final Hamiltonians in the Heisenberg picture as 
$H_0=\sum_n \epsilon_n^0 \Pi_n^0$ and $H_1=\sum_n \epsilon_n^1 \Pi_n^1$, 
where $\Pi_n^0,\Pi_n^1$ are projective operators corresponding to eigenvalues $\epsilon_n^0,\epsilon_n^1$. 

Under the stochastic description, work performed in this process is a random variable, denoted by $W$. 
The corresponding probability distribution function is $P(w)$, which is related to the probability by 
$\Pr(a<W \leq b)=\int_a^bP(w)\ud w$. 
If the initial state does not contain quantum coherence, that is, $[\rho,H_0]=0$, 
the standard two-projective measurement (TPM) gives the following quantum work probability distribution: 
\begin{equation}\label{TPM}
    \begin{aligned}
    P_{\mathrm{TPM}}(w)&= \sum_{ij}  p_{ji}p_i\delta(w-(\epsilon_j^1-\epsilon_i^0))\\
    &= \sum_{ij}  \Tr[\Pi_j^1\Pi_i^0\rho\Pi_i^0\Pi_j^1] \delta(w-(\epsilon_j^1-\epsilon_i^0)).
    \end{aligned}
\end{equation}
The above expression has a clear physical meaning. 
$p_i=\Tr[\Pi_i^0\rho\Pi_i^0]$ is the probability of the $i$-th energy level when measuring the initial Hamiltonian. 
Supposing $\tilde\rho_i=\Pi_i^0\rho\Pi_i^0/\Tr[\Pi_i^0\rho\Pi_i^0]$ is the collapsed state after the first measurement, 
$p_{ji}=\Tr[\Pi_j^1\tilde\rho_i\Pi_j^1]$ is the transition probability from the $i$-th initial energy level to the $j$-th final energy level 
while doing the second measurement after the time evolution. 
Since the system is isolated, 
the work is identified as the energy difference $\epsilon_j^1-\epsilon_i^0$ between two measurements with the associated probability $p_{ji}p_i$ of this realization. 

For the states without initial coherence, the first measurement does not perturb the overall quantum state. 
In this case, the TPM work distribution is consistent with the first law of thermodynamics as well as work fluctuation theorems. 
However, in the presence of initial quantum coherence, the first measurement destroys the initial state, and 
the off-diagonal terms of the density matrix in the energy basis are eliminated. 
Due to its invasive nature, the TPM scheme for quantum work fails to describe a process with initial coherence. 
As a direct consequence, the average work defined by the TPM scheme is in contradiction with the first law of thermodynamics. 
Despite the success of quantum thermodynamics based on the TPM scheme, 
its validity is restricted to only a small portion of initial states. 

Furthermore, a famous no-go theorem shows that under a genuine probability framework, 
there is no way to generalize the TPM scheme to incorporate initial coherence \cite{nogo}. 
In Ref. \cite{nogo}, a genuine probability distribution implies the following three constraints on $P(w)$. 
The first two constraints come from the axioms of probability theory and are easy to understand: 
$P(w)\geq 0$, and $\int_{-\infty}^{+\infty} P(w)\ud w=1$. 
The third constraint is a property regarding the initial state known as convexity: 
For any $0\leq\lambda\leq 1$, the distribution satisfies 
$P(w;\lambda \rho_1+ (1-\lambda)\rho_2)=\lambda P(w;\rho_1)+(1-\lambda)P(w; \rho_2)$. 
An explanation of convexity is given in the next section. 
The no-go theorem for quantum work \cite{nogo} states that 
the following two requirements cannot be simultaneously fulfilled under the above genuine probability framework. 
One is the first law of thermodynamics for the average work, $ \ev W = \Tr[(H_1-H_0)\rho] $. 
The other is the reduction to the standard TPM result for initial states without coherence. 
These two requirements are both very important in quantum thermodynamics but contradict each other. 

\section{requirements for quasiprobability of quantum work}\label{s_requirements}

The quasiprobability approach abandons the positivity assumption of the probability distribution function 
to resolve the contradiction. 
In the quasiprobability framework, 
the distribution function $P(w)$ could take negative or even complex values. 
Other two constraints, namely, the normalization relation and the convexity still hold. 
We summarize here the modified conditions for a quasiprobability distribution. 
\begin{itemize}
\item[(W0)] Distribution function is complex-valued (or real-valued) and normalized 
\begin{equation}
    \begin{aligned}
    P(w) \in \mathbb{C} ,\\
    \int P(w)\ud w =1.
    \end{aligned}
\end{equation}
\item[(W1)] The distribution function exhibits convexity with respect to the initial density operator: 
For any $0\leq \lambda\leq 1$, 
\begin{equation}\label{convexity}
    P(w;\lambda \rho_1+ (1-\lambda)\rho_2)=\lambda P(w;\rho_1)+(1-\lambda)P(w; \rho_2) .
\end{equation}
\end{itemize}
This convexity condition originates from the nature of the classical probability because 
the mixed density matrix $\lambda \rho_1+ (1-\lambda)\rho_2$ is equivalent to randomly choosing the initial state from $\rho_1$ 
and $\rho_2$ with (classical) probability $\lambda$ and $1-\lambda$ respectively, and 
the result for $\lambda \rho_1+ (1-\lambda)\rho_2$ should be a direct combination of two independent parts.

Since the average work in quantum mechanics is well-defined and has been widely used for a long time before the concept of stochastic work, 
there is no doubt that the first law of thermodynamics at the ensemble average level should be a criterion. 
\begin{itemize}
    \item[(W2)] The first law of thermodynamics for average work 
    \begin{equation}\label{first_law}
        \int w P(w)\ud w=\Tr\left[(H_1-H_0)\rho\right] .
    \end{equation}
\end{itemize}
Physically, the first law requires the average work to be the average of the energy change. 

Another condition that appears in the formulation of no-go theorem for quantum work is the reproduction of TPM scheme, 
which requires the quantum work to match the standard definition by TPM scheme in the case when the initial state contains no quantum coherence. 
The achievements of quantum thermodynamics such as the quantum fluctuation theorems based on the TPM scheme serve as a justification. 
Nevertheless, this requirement is not directly used in the later proof but appears as a consequence of other requirements. 
We regard it as an extra condition which is labeled as requirement (E1) in this paper. 
\begin{itemize}
    \item[(E1)] Reproduction of TPM scheme: 
    \begin{equation}\label{reduction}
        P(w)=P_{\mathrm{TPM}}(w), \quad \text{if } \comm{\rho}{H_0}=0 .
    \end{equation} 
\end{itemize}

Before listing other requirements, it is meaningful to introduce the consequences of the above requirements. 
The convexity shows the distribution exhibits a linear dependence on the initial state, 
and we assert that $P(w)$ can be expressed as 
\begin{equation}\label{r_con}
    P(w)=\Tr[M(w)\rho] ,
\end{equation}
where $M(w)$ is a bounded operator irrelevant to $\rho$. 
This expression is intuitively valid because the right-hand side also keeps the linear property on $\rho$. 
Strict proof is given in Appendix \ref{a_a} based on functional analysis. 

As illustrations, we summarize the expressions of $M(w)$ in some typical definitions for quantum work. 
From Eq. \eqref{TPM}, $M(w)$ for the TPM work distribution is 
\begin{equation}
    M_{\mathrm {TPM}}(w)=\sum_{ij}\Pi_i^0\Pi_j^1\Pi_i^0 \delta(w-(\epsilon^1_j-\epsilon^0_i)) .
\end{equation}
A $q$-class of quasiprobability distributions proposed in Refs. \cite{family,most_general} generalizes 
the MH work quasiprobability distribution and FC work quasiprobability distribution. 
$M(w)$ for $q$-class is 
\begin{equation}\label{q-class}
    \begin{aligned}
    M_q(w)=\sum_{ijk}\frac{1}2\left[\Pi_k^0 \Pi_j^1 \Pi_i^0+\Pi_i^0\Pi_j^1\Pi_k^0\right]\\
    \times\delta(w-[\epsilon_j^1-q\epsilon_i^0-(1-q)\epsilon_k^0]) ,
    \end{aligned}
\end{equation}
with a parameter $0\leq q\leq 1/2$. 
The cases with $q=0$ and $q=1/2$ reduce to the MH and FC work quasiprobability distributions respectively. 

For initial states without coherence, $\comm{\rho}{\Pi^0_n}=0$, 
all the above distribution functions coincide with each other. 
On the other hand, we can easily verify the first law for the quasiprobability in $q$-class. 
Thus, in contrast to the no-go theorem, as we abandon the positivity of probability, we arrive at a multiple-choice result: 
there are many work quasiprobability distributions that satisfy both the first law (W2) and the reproduction of the TPM scheme (E1). 

In addition to the above conditions (W0)-(W2) and (E1), 
from physical consideration, we list several other requirements. 
They are later used to demonstrate the validity of the MH quasiprobability of work. 

The following requirement plays the core role in this paper, which roots in our physical understanding of the concept of work. 
\begin{itemize}
    \item [(W3)] Support condition: The distribution function $P(w)$ is solely determined by $H_0$, $H_1$, and $\rho$. 
    If $P(w)\neq 0$, then there exists $i,j$ such that $w= \epsilon^1_j-\epsilon_i^0$. 
    And if $k\neq i,l\neq j$ for all possible $i,j$, then $P(w)$ does not depend on $\epsilon^0_k,\Pi^0_k$ and $\epsilon^1_l,\Pi^1_l$.
\end{itemize}
Here is an explanation of this requirement. 
As is shown in most proposals, the quantum work is viewed as a quantity 
related to the initial and final Hamiltonians in the Heisenberg picture, $H_0$ and $H_1$, 
with no regard to the evolution details during the process. 
We adopt this as an assumption for an isolated quantum process; work only depends on $H_0$, $H_1$, and the initial state $\rho$. 
The distribution function can be explicitly written as $P(w;H_1,H_0,\rho)$. 
Furthermore, inspired by the physical meaning of the TPM scheme, we assume that 
the stochastic work characterizes the energy level transition. 
Since the energy is distributed only at the eigenvalues, 
the support (nonzero points) of $P(w)$ is restricted to the differences between the final and initial energy eigenvalues. 
That is, $P(w)\neq 0$ implies $w$ can be expressed as $\epsilon_{j}^1-\epsilon_{i}^0$. 
Also, we expect that the transition probability between two energy levels has nothing to do with other energy levels. 
In consideration of possible degeneracy of the energy difference, let us suppose $w$ belongs to the support of $P(w)$, and 
$(i_1,j_1),(i_2,j_2),\dots$ are all the pairs such that $w=\epsilon_{j_m}^1-\epsilon_{i_m}^0$. 
If $k\neq i_m$ for any $m$, then $P(w)$ has nothing to do with the $k$-th initial energy level, 
and $P(w)$ does not depend on $\epsilon_k$ and $\Pi^0_k$. 
This is also true for the $l$-th final energy level if $l\neq j_m$ for any $m$. 
This accounts for the last sentence in (W3). 

In this paper, we do not directly require the reproduction of the TPM scheme (E1) 
but try to extract the physical implication behind the TPM scheme and introduce the support condition (W3). 
As we will see, the reproduction of the TPM scheme will emerge as a consequence of other requirements. 

Furthermore, 
there should be a linear relation between work and Hamiltonian, which was explicitly proposed in Ref. \cite{most_general}. 
Let us consider a global rescaling transform of the time-dependent Hamiltonian, $H_H^\prime(t)=\lambda H_H(t)$. 
While $\rho$ is fixed, we expect $W^\prime$ in this rescaled protocol to exhibit the same distribution as $\lambda W$. 
In another transform, the Hamiltonian is shifted by a c-number (classical number) $H_H^\prime(t)= H_H(t)+e(t)$, 
which corresponds to different choices of the zero point of the energy. 
In this case, we naturally expect $W^\prime$ to exhibit the same distribution as $W+e(\tau)-e(0)$. 
Combining these two kinds of transforms, we have the following requirement. 
\begin{itemize}
\item[(W4)] A linear rescaling condition on the Hamiltonian: 
Under the rescaling transform $H_H(t)\rightarrow  \lambda H_H(t) +e(t)$ with fixed initial state, 
the work rescales as $W\rightarrow W^\prime \overset{d}{=}\lambda W +e(t)-e(0)$. 
Here, $\overset d= $ means both sides have the same distribution. 
\end{itemize}

The zeroth-order moment and the first-order moment are fixed by the normalization relation and the first law, respectively. 
Besides, in stochastic thermodynamics, the amplitude of fluctuation can be quantified by the variance,
which should be always non-negative. Equivalently, the second-order moment is also non-negative. 
We adopt this as a requirement of work distribution for arbitrary initial states. 
\begin{itemize}
    \item [(W5)] Positivity of the second-order moment: 
    \begin{equation}
        \ev{W^2}=\int w^2 P(w)\ud w\geq 0 .
    \end{equation}
\end{itemize}

The time-reversal symmetry is the last requirement for $P(w)$, which was proposed in Ref. \cite{Miller_2017}. 
In the standard formulation, the time-reversed process of an original process is implemented as follows. 
The anti-unitary time-reversal operator $\Theta $ is involved, and we will use an overline to denote quantities in the time-reversed process. 
In the Schrödinger picture, 
the initial state is set to be the time-reversed final state in the original process $\bar\rho(0)=\Theta \rho(\tau)\Theta^{-1}$. 
The time-reversed Hamiltonian is also equipped with the time-reversal operator, and the time parameter in the Hamiltonian evolves backward: 
$\bar H_S(t)=\Theta H_S(\tau-t) \Theta^{-1}$. 
In the time-reversed process, 
thermodynamic quantities including work should exhibit a minus sign compared to the original process according to the time-reversal symmetry. 
\begin{itemize}
\item[(E2)] Time-reversal symmetry for work: 
In the time-reversed process, the corresponding work $\bar W$ has the same distribution as the minus work $-W$ in the original process: 
\begin{equation}
    \bar P(w)=P(-w) .
\end{equation}
\end{itemize}
The time-reversal symmetry plays a crucial role in stochastic thermodynamics, 
especially in the formulation of fluctuation theorems \cite{FTs}. 
Requirement (E2) is indispensable for an appropriate work quasiprobability. 
In the proof in the next section, it will not be directly used but appear as the consequence of other requirements, 
so we label it as (E2).

\section{validity of Margenau-Hill quasiprobability of work}\label{proof}
After the above preparations, we arrive at 
\textit{our main result}: In an isolated driving process, the MH quasiprobability of work is the only one 
satisfying all the requirements (W0)-(W5) among all the possible definitions. 
Meanwhile, the MH quasiprobability of work also satisfies other two requirements (E1)-(E2). 

\textit{Sketch of the proof}: 
We denote the distribution function as $P(w;H_1,H_0,\rho)$. 
According to the corollary of the convexity condition (shown in Eq. \eqref{r_con}), 
we have proved that the distribution function can be expressed as 
$P(w;H_1,H_0,\rho)=\Tr[M(w; H_1,H_0)\rho]$. 
The proof of Eq. \eqref{r_con} can be found in Appendix \ref{a_a}. 

Due to the support condition (W3), $P(w)$ only distributes at $\epsilon_j^1-\epsilon_i^0$, so
\begin{equation}
    M(w; H_1,H_0)=\sum_{ij}\delta(w-(\epsilon_j^1-\epsilon_i^0)) N_{ji} .
\end{equation}
Here the operator $N_{ji}$ is solely dependent on $H_1,H_0$, which is to be determined in Eq. \eqref{expansion_1}. 
As operators are fully determined by their eigenvalues and corresponding projectors, we can regard 
$N_{ji}$ as a function of all the eigenvalues and projectors, namely $\{\epsilon^1_k\},\{\epsilon^0_k\},\{\Pi_k^1\},\{\Pi_k^0\}$. 
Since the above requirements should be satisfied by all kinds of processes, for simplicity, we assume the energy difference is non-degenerate, 
that is, $\epsilon_j^1-\epsilon_i^0$ are different for different pairs of $(i,j)$. 
Again from (W3), only the relevant energy levels and corresponding $\epsilon_j^1,\epsilon_i^0,\Pi_j^1,\Pi_i^0$ 
contribute to the work distribution when considering $N_{ji}$. 
According to the rescaling requirement (W4), when the eigenvalues $\epsilon_j^1,\epsilon_i^0$ are shifted, 
the amplitude of the distribution function remains invariant. 
Therefore, $N_{ji}$ should not depend on eigenvalues, and eventually, $N_{ji}(\Pi_j^1,\Pi_i^0)$ is a function of two projective operators. 
We assume it to be an analytical function, which can be expanded as the following series with coefficients $e_0,a_n,b_n,c_n,d_n$: 
\begin{equation}\label{expansion_1}
    \begin{aligned}
    N_{ji}(\Pi_j^1,\Pi_i^0)=e_0+\sum_{n=1}^\infty a_n (\Pi_j^1\Pi_i^0)^n+\sum_{n=1}^\infty b_n(\Pi_i^0\Pi_j^1)^n \\
    +\sum_{n=0}^\infty c_n (\Pi_i^0\Pi_j^1)^n\Pi_i^0+\sum_{n=0}^\infty d_n\Pi_j^1(\Pi_i^0\Pi_j^1)^n .
    \end{aligned}
\end{equation}
In each term of the expansion, two different projective operators appear alternatively.

Next, we compare the average of the work to the result from the first law of thermodynamics (W2), 
\begin{equation}
    \ev{W}=\Tr[\rho \sum_{ij}(\epsilon_j^1-\epsilon_i^0)N_{ji}]=\Tr[\rho(H_1-H_0)] .
\end{equation}
Since the first law is valid for any state $\rho$, we get the following relation: 
\begin{equation}\label{con_first}
    \sum_{ij}(\epsilon_j^1-\epsilon_i^0)N_{ji}=H_1-H_0 .
\end{equation}
The next step is to use the above condition to determine the coefficients in the expansion in Eq. \eqref{expansion_1}. 
This is accomplished by constructing a specific set of projective operators. 
The detailed treatment is included in the Appendix \ref{a_b}. 
The result shows that the remaining possible choice for $N_{ji}$ is 
\begin{equation}\label{r_N}
    N_{ji}=\frac{1+\eta}2 \Pi_j^1\Pi_i^0+\frac{1-\eta}2 \Pi_i^0\Pi_j^1 .
\end{equation}
Here $\eta$ is an arbitrary complex number $\eta$. 
Also, in the Appendix \ref{a_b}, we show that the quasiprobability of work defined by the above $N_{ji}$ 
satisfies the time-reversal symmetry (E2), regardless of the value of the complex number $\eta$. 

In the last step of the proof, we will show that only when $\eta=0$, the second-order moment is positive. 
Direct calculation yields 
\begin{equation}
    \begin{aligned}
    &\ev{W^2}=\int w^2P(w)\ud w\\
    &=\Tr[(H_1-H_0)^2\rho]-\eta\Tr[(H_1H_0-H_0H_1)\rho] .
    \end{aligned}
\end{equation}
If $\eta=0$, $\ev{W^2_{\text{MH}}}=\Tr[(H_1-H_0)^2\rho]$ is always non-negative. 

For any $\eta\neq 0$, we suppose $H_1=H_0+L$, where $L$ is a Hermitian operator characterizing the difference. 
The expression for $\ev{W^2}$ changes to 
\begin{equation}\label{secon_order}
    \ev{W^2}=\Tr(L^2\rho)-\eta\Tr(\comm{L}{H_0}\rho) .
\end{equation}
The positivity of the second-order moment should hold for an arbitrary protocol, 
so we have the freedom to select the value of $L,\rho$. 
The first term depends on $L$ quadratically while the second term linearly. 
Therefore, as long as $\eta\neq 0$, for sufficiently small $L$, the second term will be dominant, which is not positive definite. 
As a result, $\ev{W^2}$ cannot always be positive if $\eta\neq 0$. 

As an explicit example, $H_0=\sigma_z$, $L=\nu\sigma_x$, $\comm{L}{H_0}=-2\nu\ui \sigma_y$, where 
$\sigma_i$ are Pauli matrices and $0<\nu\ll \eta$ is a tiny number. 
By choosing density operators $\rho=\frac 12 (I\pm \sigma_y)$, we can make
\begin{equation}
    -\eta\Tr[\comm{L}{H_0}\rho]=\pm 2\ui\eta\nu .
\end{equation}
Since the term $\Tr[L^2\rho]=\nu^2$ is negligible, 
$\ev{W^2}$ can be either complex if $\operatorname{Re} \eta\neq 0$ or negative if $\operatorname{Re} \eta = 0$.

In conclusion, the following distribution function corresponding to $\eta=0$ is uniquely determined, 
\begin{equation}
    P_{\mathrm{MH}}(w)=\frac 12 \sum_{j,i}\Tr[(\Pi_j^1\Pi_i^0+ \Pi_i^0\Pi_j^1)\rho] \delta(w-(\epsilon_j^1-\epsilon_i^0)).
\end{equation}
This is the familiar Margenau-Hill work quasiprobability \cite{w_MH}. 
For initial states without coherence, it reduces to the standard TPM scheme. 
This manifests that the requirements in this paper also justify the TPM scheme in the absence of initial coherence. 

\section{comparison of several definitions}\label{discussion}
In Table \ref{list}, we compare different definitions of quantum work and test the requirements for each of them. 
As is shown in the table, the MH work quasiprobability is the only definition 
fulfilling all the requirements. 

\begin{table*}[htbp]
    \begin{ruledtabular}
        \begin{tabular}{l|ccccccc}
            &
             \begin{tabular}[c]{@{}c@{}}Value\\ (W0)\end{tabular} &
             \begin{tabular}[c]{@{}c@{}}Linearity on\\ $\rho,H(t)$\footnote{In this condition, the convexity on initial state (W1) and the linearity with respect to the Hamiltonian (W4) are combined.}\\ (W1) (W4)\end{tabular} &
             \begin{tabular}[c]{@{}c@{}}First law\\ (W2)\end{tabular} &
             \begin{tabular}[c]{@{}c@{}}Support\\ condition\\ (W3)\end{tabular} &
             \begin{tabular}[c]{@{}c@{}}Reproduce\\ TPM result\\ (E1)\end{tabular} &
             \begin{tabular}[c]{@{}c@{}}Time-reversal\\ symmetry\\ (E2)\end{tabular} &
             \begin{tabular}[c]{@{}c@{}}Positivity of\\ second moment\\ (W5)\end{tabular} \\ \hline
           Kirkwood-Dirac \cite{w_SE,w_KD}        & Complex      & $\checkmark$ & $\checkmark$ & $\checkmark$ & $\checkmark$ & $\checkmark$ & $\times$     \\
           Margenau-Hill \cite{w_MH}        & Real\footnote{``Real" implies possibly negative.}         & $\checkmark$ & $\checkmark$ & $\checkmark$ & $\checkmark$ & $\checkmark$ & $\checkmark$ \\
           Full-Counting \cite{w_FCS}        & Real         & $\checkmark$ & $\checkmark$ & $\times$     & $\checkmark$ & $\times$     & $\checkmark$ \\
           $q$-Class($0<q<1/2$) \cite{family}            & Real         & $\checkmark$ & $\checkmark$ & $\times$     & $\checkmark$ & $\times$     & $\checkmark$ \\
           Consistent Histories \cite{Miller_2017} & Real         & $\checkmark$ & $\checkmark$ & $\times$     & $\times$     & $\checkmark$ & $\checkmark$ \\
           Operator of work \cite{o-of-work_77,o-of-work_05}     & Non-negative & $\checkmark$ & $\checkmark$ & $\times$     & $\times$     & $\checkmark$ & $\checkmark$ \\
           TPM \cite{TPM1,TPM2}                 & Non-negative & $\checkmark$ & $\times$     & $\checkmark$ & $\checkmark$ & $\times$     & $\checkmark$ \\
           Hamilton-Jacobi \cite{w_Bohm}\footnote{This definition is also known as quantum work in the Bohmian framework.}      & Non-negative & $\times$     & $\checkmark$ & $\times$     & $\times$     & $\checkmark$ & $\checkmark$
           \end{tabular}
    \end{ruledtabular}
    \caption{Comparison of different definitions of work distributions: $\checkmark$ is used only if the requirement is fulfilled under all the possible processes. }
    \label{list}
\end{table*}

Definitions with positive distribution face severe flaws. 
In accordance with the no-go theorem, at least one of the following conditions is violated: the first law, the reproduction of TPM, or the convexity. 
For detailed review of positive-valued work probabilities including Quantum Hamilton-Jacobi \cite{w_Bohm}, Gaussian measurements \cite{PhysRevE.93.022131}, 
POVM depending on the initial state \cite{sagawa2013}, please refer to Ref. \cite{book_QW}. 

In what follows, we will mainly focus on quasiprobability distributions. 
According to Eq. \eqref{q-class}, the quasiprobability distributions in $q$-class ($0<q\leq 1/2$, including FC) demonstrate an asymmetry between the initial and the final Hamiltonians. 
The initial projector appears twice but the final projector only once. 
Therefore, the time-reversal symmetry is violated unless $q=1/2$. 
Besides, the $q$-class $(0< q\leq 1/2)$ does not satisfy the support condition. 
The work distributes at $\epsilon^1_j-[q\epsilon^0_i+(1-q)\epsilon^0_k]$, where the initial energy undergoes a weighted average of two initial eigenvalues. 

In contrast, the Consistent-History (CH) quasiprobability of work demonstrates time-reversal symmetry \cite{Miller_2017}. 
Unlike most definitions, CH work involves the evolution details. 
In this formulation, a trajectory is built based on the power operator $\partial_t H_H(t)$, and 
the work is associated with the integrated power during the whole process. 
The reproduction of the TPM result (E1) and the fluctuation theorems cannot be fulfilled 
because work is not a quantity related to the initial and final Hamiltonians in this proposal. 
By the way, we remark that usually, if (E1) is violated, then so is the support condition (W3). 
This justifies (W3) as a more strict generalization of (E1). 

It is noticeable that the Kirkwood-Dirac (KD) quasiprobability of work also satisfies all the requirements, 
apart from its complex value as well as complex second moment. 
Its distribution function is given by 
\begin{equation*}
    P_{\mathrm{KD}}(w)=\sum_{ij}\Tr[\Pi^1_j\Pi^0_i\rho]\delta(w-(\epsilon^1_j-\epsilon^0_i)).
\end{equation*}
Actually, the MH distribution is just the real part of the KD distribution, 
and it is wise to first calculate the KD distribution and take the real part to obtain the MH one in many cases. 
\begin{equation}
    P_{\mathrm{MH}}(w)=\operatorname{Re} P_{\mathrm{KD}}(w) .
\end{equation}
Likewise, if we loosen the positivity of the second moment (W5) to the positivity of the real part of the second moment, 
the following KD quasiprobability with a rescaled imaginary part meets all the other requirements. 
\begin{equation}
    P_\xi(w)=\operatorname{Re} P_{\mathrm{KD}}(w) +\ui \xi\operatorname{Im} P_{\mathrm{KD}}(w),
\end{equation}
with a real parameter $\xi$. 
Since the $\operatorname{Im} P_{\mathrm{KD}}(w)$ can be rescaled without violating most requirements, 
the physical meaning of the imaginary part in the KD distribution is not clear. 
We suggest that the real part and imaginary part of the KD distribution should be treated separately because of their distinct behaviors. 
Fortunately, the strict requirement on the positivity of secondary moment (W5) rules out these complex-valued candidates, 
leaving the MH quasiprobability of quantum work the only appropriate definition.

\section{Example: Breathing harmonic oscillator with initial squeezed states}\label{s_example}

We adopt the MH quasiprobability and evaluate the
work distribution for the breathing harmonic oscillator. 
We will further verify the quantum-classical correspondence principle of work distribution 
for initial states with quantum coherence in this example \cite{Pan2019}. 
The Hamiltonian in the Schrödinger picture is 
\begin{equation}
\hat{H}_S(t)=\frac{\hat{p}^{2}}{2m}+\frac{1}{2}m\omega(t)^{2}\hat{x}^{2}.
\end{equation}
The frequency $\omega(t)$ is varied in a finite-time process. 
In this section, we use the symbol `` $\hat{}$ " to denote operators in quantum mechanics. 

Let us adopt a special protocol to vary the frequency \cite{e18050168,otto_1,PhysRevE.103.022136}. 
\begin{equation}
\omega(t)=\frac{\omega_{0}}{(\omega_{0}/\omega_{\tau}-1)t/\tau+1},
\end{equation}
where $\omega(0)=\omega_0$, and $\omega(\tau)=\omega_\tau$. 
The final Hamiltonian in the Heisenberg picture 
can be obtained analytically as \cite{PhysRevA.105.022609} 
\begin{equation}\label{Htau}
    \begin{aligned}
    \hat{H}_{H}(\tau)&=\frac{1}{2m}\hat p_H(t)^2+\frac 12 m\omega(t)^2\hat x_H(t)^2\\
    &=\frac{\Gamma_{11}}{2}\hat{x}^{2}+\frac{\Gamma_{22}}{2}\hat{p}^{2}  +\frac{\Gamma_{12}}{2}(\hat{x}\hat{p}+\hat{p}\hat{x}) .
    \end{aligned}
\end{equation}
The coefficients are 
\small
\begin{equation*}
\begin{aligned}
    \Gamma_{11} & =\ue^{r}\left[\frac{\cosh(\sqrt{1-g^{2}}r)-g^{2}}{1-g^{2}}+\frac{\sinh(\sqrt{1-g^{2}}r)}{\sqrt{1-g^{2}}}\right]m\omega_0^2,\\
    \Gamma_{22} & =\ue^{r}\left[\frac{\cosh(\sqrt{1-g^{2}}r)-g^{2}}{1-g^{2}}-\frac{\sinh(\sqrt{1-g^{2}}r)}{\sqrt{1-g^{2}}}\right]/m,\\
    \Gamma_{12} & =\ue^{r}\frac{g\left[\cosh(\sqrt{1-g^{2}}r)-1\right]}{1-g^{2}}\omega_0,
\end{aligned}
\end{equation*}
\normalsize
where $g=2\omega_{0}\omega_{\tau}\tau/(\omega_{\tau}-\omega_{0})$, and $r=\ln(\omega_{\tau}/\omega_{0})$.

In the following, we consider an initial squeezed state in the form 
\begin{equation}\label{init}
    \begin{aligned}
    \hat{\rho} =&2\sinh(\frac{1}{2}\hbar\sqrt{\gamma_{11}\gamma_{22}-\gamma_{12}^{2}})\\
    \times&\exp(-\frac{1}{2}[\gamma_{11}\hat{x}^{2}+\gamma_{12}(\hat{x}\hat{p}+\hat{p}\hat{x})+\gamma_{22}\hat{p}^{2}]).
    \end{aligned}
\end{equation}
Here $\gamma_{11}$, $\gamma_{22}$, and $\gamma_{12}$ are three parameters characterizing the squeezed state. 
To evaluate the work distribution, we first calculate the characteristic function of the quantum work under the 
MH quasiprobability 
\begin{equation}
    \begin{aligned}
    \chi_{W}(s)&=\int\ue^{\ui sw}P_{\text{MH}}(w)\ud w\\
    &=\mathrm{Tr}\{\frac{1}{2}[\ue^{\ui s\hat{H}_{H}(\tau)}\ue^{-\ui s\hat{H}(0)}+\ue^{-\ui s\hat{H}(0)}\ue^{\ui s\hat{H}_{H}(\tau)}]\hat{\rho} \}.
    \end{aligned}
\end{equation}
It can be evaluated by the trace formula of a single bosonic mode \citep{PhysRevResearch.1.033175}
\begin{equation}
    \Tr\left[\prod_{j}\ue^{\frac{1}{2}R^{T}M_{j}R}\right]=\left\{ -\det\left[\prod_{j}\ue^{-\hbar\sigma_{y}M_{j}}-I\right]\right\} ^{-\frac{1}{2}},
\end{equation}
where the position and the momentum are given by $R=(\begin{array}{cc} 
\hat{x} & \hat{p}\end{array})^{T}$, $\sigma_{y}$ is the second Pauli matrix,  $I$ is identity $2\times2$ matrix, and $M_{j}$ ($j=1,2,3$) are symmetric 
$2\times2$ matrices
\begin{align}
    M_{1} & =\ui s\left(\begin{array}{cc}
    \Gamma_{11} & \Gamma_{12}\\
    \Gamma_{12} & \Gamma_{22}
    \end{array}\right),\\
    M_{2} & =-\ui s\left(\begin{array}{cc}
    m\omega_0^2\\
     & 1/m
    \end{array}\right),\\
    M_{3} & =-\left(\begin{array}{cc}
        \gamma_{11} & \gamma_{12}\\
        \gamma_{12} & \gamma_{22}
    \end{array}\right).
\end{align}
The corresponding work distribution $P_{\text{MH}}(w)$ is obtained as the inverse Fourier transform, 
and it appears to be a discrete distribution. 

On the other hand, the classical work distribution can be obtained from the initial distribution in phase space, 
\begin{equation}
    \rho(x,p)=\frac{\sqrt{\gamma_{11}\gamma_{22}-\gamma_{12}^{2}}}{2\pi}\ue^{-\frac{1}{2}(\gamma_{11}x^{2}+2\gamma_{12}xp+\gamma_{22}p^{2})},
\end{equation}
which represents the classical counterpart of the quantum state given in Eq. \eqref{init}. 
The work distribution is obtained as the integral in the phase space 
\begin{equation}\label{eq:P_cl(w)}
    P^{\mathrm{cl}}(w)=\iint\rho(x,p)\delta(w-H_{H}(\tau)+H(0))\ud x\ud p.
\end{equation}
Here, $H_H(\tau)$ is also given by Eq. \eqref{Htau}, but it is a c-number determined by the initial point in phase space. 
The result of the characteristic function is 
\small
\begin{equation*}
    \begin{aligned}
    &\chi_{W}^{\mathrm{cl}}(s)=\iint \ue^{\ui s(H_{H}(\tau)-H(0))}\rho(x,p)\ud x\ud p=\\
    &\sqrt{\frac{\gamma_{11}\gamma_{22}-\gamma_{12}^{2}}{[\gamma_{11}-\ui s(\Gamma_{11}-m\omega_0^2)][\gamma_{22}-\ui s(\Gamma_{22}-\frac 1m)]-(\gamma_{12}-\ui s\Gamma_{12})^{2}}}.
    \end{aligned}
\end{equation*}
\normalsize
The corresponding work distribution is also obtained by the inverse Fourier transform, but it is a continuous distribution.

\begin{figure}
\includegraphics[width=8cm]{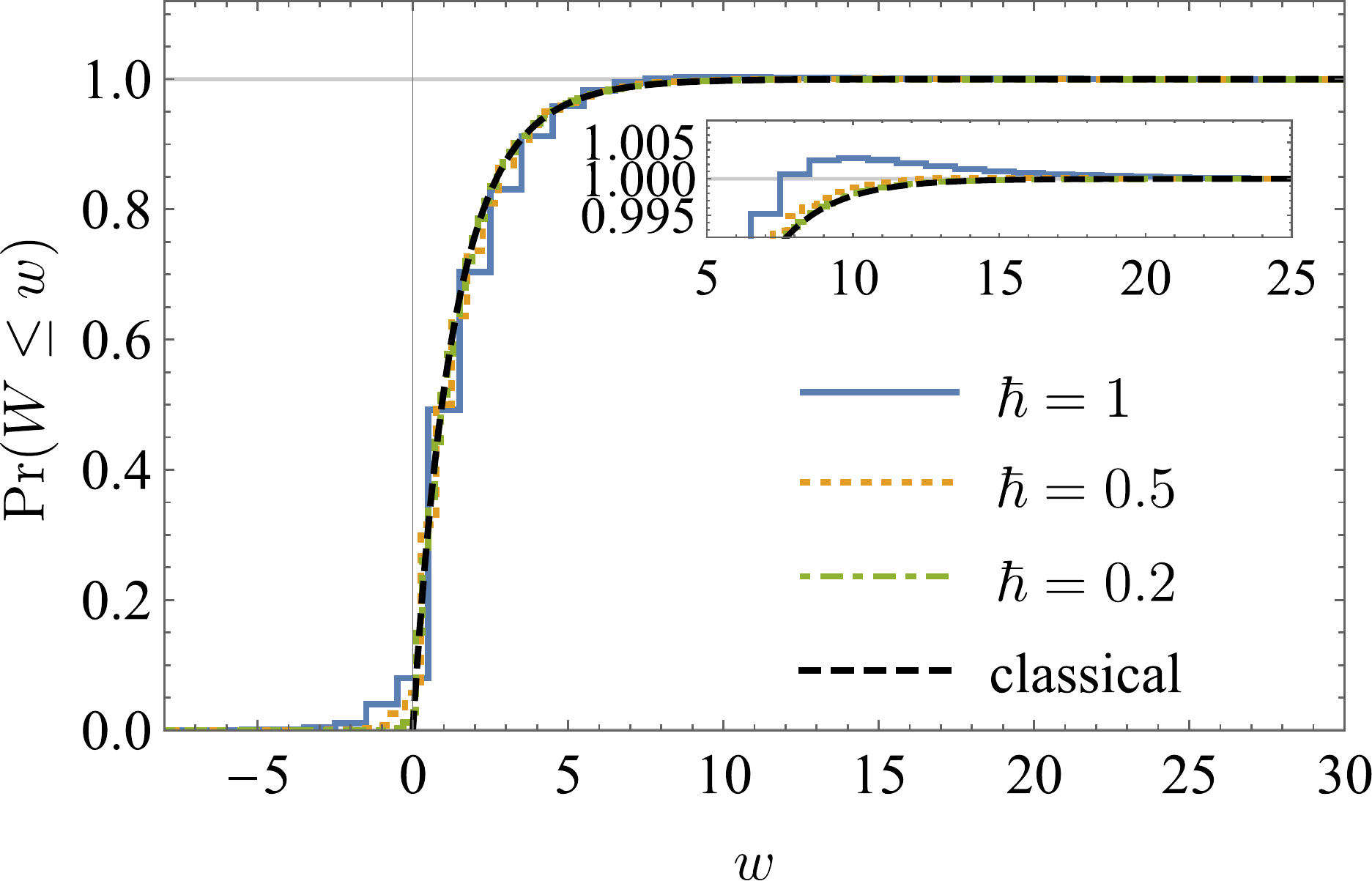}

\caption{Accumulated distribution of work $\mathrm{Pr}(W \leq w)$. The black dashed
curve corresponds to classical work distribution. The discrete distributions
with blue solid, orange dotted, and green dot-dashed lines correspond
to MH quasiprobability of quantum work with $\hbar=1,0.5,0.2$. The
initial squeezed state is prepared with $\gamma_{11}=1$, $\gamma_{22}=1.1$,
and $\gamma_{12}=0.9$. The parameters of the harmonic oscillator
are chosen as $m=1$, $\omega_{0}=1$ and $\omega_{\tau}=2$, and
the operation time is $\tau=1$. 
The inset shows the negative quasiprobability at relatively large $w$.}
\label{fig:Accumulated-distribution-of}
\end{figure}

In Fig. \ref{fig:Accumulated-distribution-of}, we show the accumulated
distribution of work $\mathrm{Pr}(W\leq w)=\int_{-\infty}^{w}P(w^{\prime})\ud w^{\prime}$
with the initial squeezed state in Eq. \eqref{init}. 
The parameters are given in the caption of the figure. 
With the decrease of $\hbar$, the discrete distributions 
of quantum work (blue solid, orange dotted, and green dot-dashed lines)
converge to the classical work distribution (black dashed curve). The convergence shows the quantum-classical 
correspondence principle of MH quasiprobability of quantum work for 
initial states with quantum coherence \cite{Pan2019}.

We remark that a generalized Jarzynski equality has
been found for arbitrary initial state \cite{w_MH,Gong2015}. In
Refs. \cite{w_MH,Pan2019}, the generalized Jarzynski equality is obtained
for initial states with quantum coherence as 
\begin{equation}\label{eq:generalized_Jarzynski_eq}
    \left\langle \ue^{-\beta W}\right\rangle =e^{-\beta\Delta F}\operatorname{Re}\Tr\left[\hat{\rho}\hat{U}^{-1}\hat{\rho}_{\mathrm{eq}}(\tau)U\hat{\rho}_{\mathrm{eq}}(0)^{-1}\right],
\end{equation}
where $\hat{\rho}_{\mathrm{eq}}(0)=\ue^{-\beta\hat{H}_{S}(0)}/\mathrm{Tr}[\ue^{-\beta\hat{H}_{S}(0)}]$ and $\hat{\rho}_{\mathrm{eq}}(\tau)=\ue^{-\beta\hat{H}_{S}(\tau)}/\mathrm{Tr}[\ue^{-\beta\hat{H}_{S}(\tau)}]$
are equilibrium states with the inverse temperature $\beta$ at the initial and the final time, and $\hat{U}$ is the evolution operator.
The conventional Jarzynski equality is recovered for an initial equilibrium
state $\hat{\rho} =\hat{\rho}_{\mathrm{eq}}(0)$. 
With the work distribution obtained in Fig. \ref{fig:Accumulated-distribution-of},
we can verify the generalized Jarzynski equality \eqref{eq:generalized_Jarzynski_eq}
by evaluating the two sides separately.

In experiments, the MH quasiprobability 
can be obtained by the Ramsey scheme \cite{Ramsey_t,exp_work,PhysRevLett.110.230602}. 
Thus, it is possible to verify the generalized Jarzynski equality
for initial states with quantum coherence in future experiments.

\section{conclusion and outlook}
In this paper, we list several requirements for an appropriate definition of work distribution. 
Building on the physical implication of the TPM scheme, the support condition is introduced, 
which reveals that the quantum work corresponds to the transition between the initial and final energy levels, 
and it plays a crucial role in our proof. 
Other requirements include the convexity with respect to the initial density operator, the first law of thermodynamics, linear relation on the Hamiltonian, 
the positivity of second-order moment, and the time-reversal symmetry. 

We prove that the MH quasiprobability of quantum work is the
only definition that satisfies all the requirements. 
We also compare some common definitions of quantum work and examine their properties from the perspective of requirements. 
The FC quasiprobability as well as the $q$-class quasiprobability fails to fulfill the time-reversal symmetry and the support condition 
due to the asymmetry between the initial and the final Hamiltonians. 
The KD quasiprobability is also an acceptable definition, which satisfies most of the requirements. 
Nevertheless, the physical meaning of its imaginary part is unexplored, 
and the appearance of the imaginary part violates the positivity requirement of the second-order moment. 
Under all the requirements in this paper, MH quasiprobability of quantum work shows its advantages over other definitions of quantum work. 

We explicitly evaluate 
the MH quasiprobability of quantum work distribution in the example of a breathing harmonic oscillator. 
The distribution exhibits slightly negative values at the tail. 
The Jarzynski equality for initial states with quantum coherence \cite{w_MH,Pan2019} is verified in this example. 

In addition to the quantum work, other thermodynamic quantities, such as heat and particle number exchange, can also be characterized by quasiprobability 
if the corresponding operators and the initial density operator do not commute with each other \cite{PRXQuantum.3.010304,YungerHalpern2016,majidy2023noncommuting}. 
The requirements in this paper can also help us justify the validity of the MH quasiprobability in such cases. 
In the study of quantum thermodynamics, one often restricts to the process without initial coherence, 
in contrast to classical stochastic thermodynamics, where no restriction is put on the process. 
With the help of the MH quasiprobability, it is promising to handle processes with initial quantum coherence. 
We will be able to calculate the full counting statistics and characterize the fluctuation for thermodynamic quantities through a standard way 
in a generic quantum nonequilibrium process. 
This will provide a solid foundation for related issues such as quantum heat engines and work extractions. 
In future studies, we hope that the quasiprobability approach can broaden the scope of quantum stochastic thermodynamics. 

\begin{acknowledgments}
    This work is supported by the National Natural Science Foundation
    of China (NSFC) under Grants No. 12147157, No. 11775001, and No. 11825501.
\end{acknowledgments}

\begin{widetext}
\appendix

\section{CONVEXITY ON INITIAL STATE AND ITS COROLLARY}\label{a_a}
In this appendix, we are going to prove that 
the convexity property (W1) of distribution implies the distribution function $P(w)$ can be expressed as
\begin{equation}\label{aim_A}
    P(w;\rho)=\Tr\left[M(w)\rho\right] ,
\end{equation}
where $M(w)$ is a bounded operator irrelevant to $\rho$. 

Before proceeding, we remark that although in the case with $P\geq 0$, this result has been proved \cite{nogo}, 
there is no strict proof if $P$ is a quasiprobability distribution with real or complex value. 
Here, we give proof that applies to Hilbert space of any (maybe infinite) dimension. 

We now briefly illustrate the outline of the proof. 
For a specific distribution, we do a linear expansion of $P(w;\rho)$ to $\tilde P(w;\tilde \rho)$, 
allowing $\tilde \rho$ to be any operator that has a definite trace, called trace-class operator. 
In the trace class, operators are no longer Hermitian or positive. 
The expanded distribution $\tilde P$ becomes a linear function in the trace-class space. 
Discussions about trace-class space can be found in textbook \cite{Conway_6}. 
The general form of a linear function of the trace-class operator $\tilde\rho$ is given by $\Tr[M\tilde \rho]$ with $M$ a bounded operator. 
Therefore, $\tilde P$ is represented in this general form, and so is $P$ when restricting to the density operators. 
In the following, we elaborate on how it works, and the feasibility of the linear expansion will be justified. 

We start with a lemma. This lemma is used in the later construction of the linear expansion. \\
\textit{Lemma 1}: 
If a linear combination of several density operators $\rho_i$ with real coefficients $\lambda_i \in \mathbb R$ is zero, 
that is, 
$$\lambda_1\rho_1 + \cdots +\lambda_N\rho_N=0 ,$$
then $\lambda_1P(w;\rho_1)+\cdots + \lambda_NP(w;\rho_N)=0$. \\
\textit{Proof}: 
$\rho_i$ are all density operators with trace 1, so $0=\lambda_1+\cdots+\lambda_N$. 
The case that all the $\lambda_i$ equal $0$ is trivial, so 
we suppose at least one element in $\{\lambda_i\}$ is greater than zero. 
We mark the non-negative terms as $\lambda_{i_k},1\leq k\leq M$ and the negative terms as $\lambda_{j_k},1\leq k\leq N-M$. 
We move the term with negative $\lambda_{j_k}$ to the right-hand side and rescaling the equation, 
\begin{equation}\label{lr}
    \frac{1}{\Lambda}\left( \abs{\lambda_{i_1}}\rho_{i_1}+\cdots+ \abs{\lambda_{i_M}}\rho_{i_{M}}\right) = \frac{1}{\Lambda} \left(\abs{\lambda_{j_1}}\rho_{j_1}+\cdots +\abs{\lambda_{j_{N-M}}}\rho_{j_{N-M}}\right) ,
\end{equation}
where $\Lambda=\abs{\lambda_{j_1}}+\cdots+\abs{\lambda_{j_{N-M}}}=\abs{\lambda_{i_1}}+\cdots+\abs{\lambda_{i_1}}$. 
The expression in the above equation is a density operator, which we will denote as $\rho^\prime$. 
Then we split the first term on the left-hand side from the other terms and write it as
\begin{equation*}
    \rho^\prime=\frac{\abs{\lambda_{i_1}}}{\Lambda} \rho_{i_1}+\frac{\Lambda-\abs{\lambda_{i_1}}}{\Lambda}\times \frac{1}{\Lambda-\abs{\lambda_{i_1}}}\left(\abs{\lambda_{i_2}}\rho_{i_2}+\cdots +\abs{\lambda_{i_M}}\rho_{i_{M}}\right),
\end{equation*}
where $\rho_{i_1}$ and $\frac{1}{\Lambda-\abs{\lambda_{i_1}}}\left(\abs{\lambda_{i_2}}\rho_{i_2}+\cdots +\abs{\lambda_{i_M}}\rho_{i_{M}}\right)$ are both density operators. 
From the convexity relation, we get 
\begin{equation*}
    P(w;\rho^\prime)=\frac{\abs{\lambda_{i_1}}}{\Lambda}P(w;\rho_{i_1})+\frac{\Lambda-\abs{\lambda_{i_1}}}{\Lambda}P(w;\frac{\abs{\lambda_{i_2}}\rho_{i_2}+\cdots +\abs{\lambda_{i_M}}\rho_{i_{M}}}{\Lambda-\abs{\lambda_{i_1}}}) .
\end{equation*}
Next, for density operator $\frac{\abs{\lambda_{i_2}}\rho_{i_2}+\cdots +\abs{\lambda_{i_M}}\rho_{i_{M}}}{\Lambda-\abs{\lambda_{i_1}}}$, 
we continue to split out the first term. After a similar rescaling and applying the convexity relation, we have 
\begin{equation*}
    \begin{aligned}
    &P(w;\frac{\abs{\lambda_{i_2}}\rho_{i_2}+\cdots +\abs{\lambda_{i_M}}\rho_{i_{M}}}{\Lambda-\abs{\lambda_{i_1}}})\\
    &=\frac{\abs{\lambda_{i_2}}}{\Lambda-\abs{\lambda_{i_1}}}P(w;\rho_{i_2})+\frac{\Lambda-\abs{\lambda_{i_1}}-\abs{\lambda_{i_2}}}{\Lambda-\abs{\lambda_{i_1}}}P(w;\frac{\abs{\lambda_{i_3}}\rho_{i_3}+\cdots+\abs{\lambda_{i_M}}\rho_{i_M}}{\Lambda-\abs{\lambda_{i_1}}-\abs{\lambda_{i_2}}}) .
    \end{aligned}
\end{equation*}
Repeating the steps to the remaining part, we finally split all the terms apart. The final result is simply 
\begin{equation*}
    \Lambda P(w;\rho^\prime)=\abs{\lambda_{i_1}}P(w;\rho_{i_1})+\cdots +\abs{\lambda_{i_M}}P(w;\rho_{i_M}) .
\end{equation*}
Similar procedure applies to the right-hand side of Eq. \eqref{lr}, 
\begin{equation*}
    \Lambda P(w;\rho^\prime)=\abs{\lambda_{j_1}}P(w;\rho_{j_1})+\cdots + \abs{\lambda_{j_{N-M}}}P(w;\lambda_{j_{N-M}}) .
\end{equation*}
Combining the two sides and moving the right-hand side back to the left, we complete the proof of the lemma. $\square $

Next, we are going to see that the density operators span the trace class \cite{Conway_6}, a subspace of the bounded operator space. 
The trace class is defined by: 
An operator $A$ belongs to the trace class if and only if $\Tr\abs{A}<\infty$, where $\abs{A}=(A^\dagger A)^\frac 12$.
If $A$ belongs to the trace class, $A$ is called a trace-class operator, which has definite trace $\Tr A$. 
Obviously, any linear combination of finite number of density operators is a trace-class operator. 

On the other hand, let us assume $A$ is a trace-class operator. 
Every bounded operator can be decomposed into a Hermitian operator plus an anti-Hermitian operator, so we have 
\begin{equation*}
    A=B+\ui C ,
\end{equation*}
where $B=\frac 12 (A+A^\dagger),C=\frac 1{2\ui}(A-A^\dagger)$ are Hermitian operators. 
Likewise, both $B$ and $C$ are trace-class operators. 
Furthermore, bounded Hermitian operators can be decomposed into a subtraction of two positive operators \cite{Conway_2}. 
This applies to $B$ and $C$: 
\begin{equation*}
    B= B_1 -  B_2 , C= C_1- C_2 .
\end{equation*}
By the way, if $B_1B_2=B_2B_1=0$ (or $C_1C_2=C_2C_1=0$), the decomposition is unique. 
Meanwhile, positive operators $B_1,B_2,C_1,C_2$ are all trace-class operators with finite trace. 
After a rescaling, operator $A$ is expressed as a linear combination of four density operators. 
\begin{equation}\label{decomposition}
    A=\lambda_1\rho_1-\lambda_2\rho_2+\ui \lambda_3\rho_3-\ui\lambda_4\rho_4 ,
\end{equation}
where $\rho_1=B_1/\Tr B_1,\rho_2=B_2/\Tr B_2,\rho_3=C_1/\Tr C_1,\rho_4=C_2/\Tr C_2$, and all the $\lambda_i$ are non-negative number. 
Therefore, any trace-class operator can be decomposed as the linear combination of 4 density operators (this decomposition is not unique generally), 
and all the density operators span the trace class.

For a specific distribution function $P(w;\rho)$, 
we define an expanded distribution $\tilde P$ with respect to any trace-class operator $A$
in the following way: 
\begin{equation}\label{expansion}
    \tilde P(w;A)=\lambda_1 P(w;\rho_1)-\lambda_2 P(w;\rho_2)+\ui \lambda_3 P(w;\rho_3)-\ui \lambda_4 P(w;\rho_4) ,
\end{equation}
provided the decomposition in Eq. \eqref{decomposition}. 
To serve as a good definition, it should be independent of the choice of decomposition. 
If there is another different decomposition $A=\lambda_1^\prime \rho_1^\prime -\lambda_2^\prime \rho_2^\prime +\ui \lambda_3\rho^\prime_3-\ui\lambda_4^\prime\rho_4^\prime$, 
the expanded distribution under this decomposition is 
\begin{equation}
    \tilde P^\prime (w;A)=\lambda_1^\prime P(w;\rho_1^\prime)-\lambda_2^\prime P(w;\rho_2^\prime)+\ui \lambda_3^\prime P(w;\rho_3^\prime)-\ui \lambda_4^\prime P(w;\rho_4^\prime) .
\end{equation}
In the two different decompositions, the Hermitian part and anti-Hermitian part equal each other respectively, 
that is, $\lambda_1\rho_1-\lambda_2\rho_2=\lambda_1^\prime \rho_1^\prime-\lambda_2^\prime\rho_2^\prime$ and $\lambda_3\rho_3-\lambda_4\rho_4=\lambda_3^\prime \rho_3^\prime-\lambda_4^\prime\rho_4^\prime$. 
According to lemma 1, $\lambda_1 P(\rho_1)-\lambda_2 P(\rho_2)=\lambda_1^\prime P(\rho_1^\prime)-\lambda_2^\prime P(\rho_2^\prime)$, 
$\lambda_3 P(\rho_3)-\lambda_4 P(\rho_4)=\lambda_3^\prime P(\rho_3^\prime)- \lambda_4^\prime P(\rho_4^\prime)$, 
so the above $P^\prime$ coincides with $P$ in Eq. \eqref{expansion}. 
This manifests that the linear expansion is independent of the decomposition. 
Besides, it is straightforward to verify that 
the expanded distribution exhibits the standard linear relation that for any complex number $\mu_1,\mu_2$ 
and trace-class operator $A_1,A_2$, 
\begin{equation}
    \tilde P(w;\mu_1 A_1+\mu_2 A_2)=\mu_1\tilde P(w;A_1)+\mu_2\tilde P(w;A_2) .
\end{equation}
Therefore, $\tilde P$ is a linear function on the trace class, and 
we have finished our treatment for the linear expansion. 

The dual space (the space of bounded linear function) of the trace class is isomorphic to the bounded operator space \cite{Conway_6}. 
As a linear function of $A$, the expanded distribution $\tilde P$ is represented as 
\begin{equation}
    \tilde P(w;A)=\Tr\left[M(w)A\right] .
\end{equation}
Here $M(w)$ is a bounded operator, which has a one-on-one correspondence with the distribution $\tilde P(w)$. 
If we restrict $A$ to the density matrix $\rho$, $\tilde P$ reduces to the work distribution $P$. 
In this case, we get the final result, 
\begin{equation}
    P(w;\rho)=\Tr\left[M(w)\rho\right] .
\end{equation}

\section{THE FIRST LAW DETERMINES THE EXPANSION COEFFICIENTS}\label{a_b}
We are going to specify how to determine the expansion coefficients in Eq. \eqref{expansion_1}
under the condition of the first law of thermodynamics 
so that the form of $N_{ji}(\Pi_j^1,\Pi_i^0)$ is fixed to Eq. \eqref{r_N}.
In addition, we will verify the time-reversal symmetry for the $N_{ji}$ in Eq. \eqref{r_N}. 

We have obtained the condition for $N_{ij}$ from the first law in Eq. \eqref{con_first}, which is also listed below: 
\begin{equation}
    \sum_{ij}(\epsilon_j^1-\epsilon_i^0)N_{ji}=H_1-H_0=\sum_j\epsilon_j^1\Pi_j^1-\sum_i\epsilon_i^0\Pi_i^0. 
\end{equation}
Since the first law is valid in arbitrary protocols, we have the liberty to assign the values of $\epsilon_j^1,\Pi_j^1,\epsilon_i^0,\Pi_i^0$. 
Due to the arbitrariness of $\epsilon_j^1,\epsilon_i^0$, the above equation is equivalent to 
\begin{equation*}
    \sum_i N_{ji}=\Pi_j^1,\; \sum_j N_{ji}=\Pi_i^0 .
\end{equation*}
Noticing in the MH work quasiprobability, $N_{ij}^{\textit{MH}}=\frac{1}2 (\Pi_j^1\Pi_i^0+\Pi_i^0\Pi_j^1)$ obviously satisfies the above equation, we adopt a trick here. 
We consider $\tilde N_{ij}=N_{ij}-N_{ij}^{\textit{MH}}$ instead, and 
a tilde is used for $\tilde N_{ij}$ in the same expansion as Eq. \eqref{expansion_1}. 
That is, 
\begin{equation}\label{series}
    \tilde N_{ji}(\Pi_j^1,\Pi_i^0)=\tilde e_0I+\sum_{n=1}^\infty[ \tilde a_n (\Pi_j^1\Pi_i^0)^n+\tilde b_n(\Pi_i^0\Pi_j^1)^n]+\sum_{n=0}^\infty[\tilde c_n \Pi_i^0(\Pi_j^1\Pi_i^0)^n+\tilde d_n(\Pi_j^1\Pi_i^0)^n\Pi_j^1 ] .
\end{equation}
The condition for $\tilde N_{ij}$ now changes to 
\begin{equation}\label{ding}
    \sum_i \tilde N_{ji}=0, \sum_j \tilde N_{ji}=0 .
\end{equation}
It is enough to find a specific set of $\Pi_j^1,\Pi_i^0$, under which all the $\tilde a_n, \tilde c_n$ are fixed to be our desired values. 
We will show how this works in the separable Hilbert space with dimension $2<d\leq\infty$.

For the sake of clarity, in this section, we denote the initial eigenstates as $\ket{\psi_m},1\leq m\leq d$, and the final eigenstates as $\ket{\phi_m}$, 
instead of $\ket{\psi^0_m},\ket{\psi^1_m}$. 
The projective operators corresponding to these eigenstates are 
\begin{equation*}
    \Pi_i^0=\outerproduct{\psi_i}{\psi_i}, \Pi_j^1=\outerproduct{\phi_j}{\phi_j} .
\end{equation*}
Direct calculation of each term in the expansion yields 
\begin{align*}
    (\Pi^1_j\Pi^0_i)^n=\innerproduct{\phi_j}{\psi_i}^n\innerproduct{\psi_i}{\phi_j}^{n-1}\outerproduct{\phi_j}{\psi_i},
\\
    (\Pi^0_i\Pi^1_j)^n=\innerproduct{\phi_j}{\psi_i}^{n-1}\innerproduct{\psi_i}{\phi_j}^n\outerproduct{\psi_i}{\phi_j},
\\
    \Pi^0_i(\Pi^1_j\Pi^0_i)^n=\innerproduct{\phi_j}{\psi_i}^n\innerproduct{\psi_i}{\phi_j}^n\outerproduct{\psi_i}{\psi_i},
\\
    (\Pi^1_j\Pi^0_i)^n\Pi^1_j=\innerproduct{\phi_j}{\psi_i}^n\innerproduct{\psi_i}{\phi_j}^n\outerproduct{\phi_j}{\phi_j}.
\end{align*}
For $2<d\leq\infty$, our specific set of eigenstates are chosen as follows. 
The initial energy eigenstates are selected to be the standard basis, that is, 
only the $m$-th component is nonzero, 
\begin{equation}\label{ini_eig}
    \ket{\psi_m}=(0,\dots,0,\overset{m} {1},0,\dots) .
\end{equation}
As for final energy eigenstates, the first three states are some real vectors: 
\begin{equation}\label{fin_eig}
    \begin{aligned}
        &\ket{\phi_1}=(&&\sin\theta\cos\phi,&&\sin\theta\sin\phi,&&\cos\theta,&&0,\dots&&),
    \\
        &\ket{\phi_2}=(&&\cos\theta\cos\phi,&&\cos\theta\sin\phi,&&-\sin\theta,&&0,\dots&&),
    \\
        &\ket{\phi_3}=(&&-\sin\phi,&&\cos\phi,&&0,&&0,\dots&&) .
    \end{aligned}
    \end{equation}
Other final energy eigenstates are the same as the initial ones, 
\begin{equation}
    \ket{\phi_m}=\ket{\psi_m},\quad \text{for } m>3 .
\end{equation}
Also, the whole final energy eigenstates form a set of orthogonal basis in the Hilbert space. 

Let us consider the following matrix element of the Eq. \eqref{ding}, 
\begin{equation}
    \sum_{j=1}^d \ev{\tilde N_{j1}}{\psi_{2}}=0. 
\end{equation}
Substituting Eqs. \eqref{series}, \eqref{fin_eig} in the above equation, 
we find only one kind of terms in the expansion survives besides the identity, and the explicit expression reads 
\begin{equation}
    d\tilde e_0+\sum_{n=0}^\infty \tilde d_n[(\sin\theta\cos\phi)^{2n}(\sin\theta\sin\phi)^2+(\cos\theta\cos\phi)^{2n}(\cos\theta\sin\phi)^2+(-\sin\phi)^{2n}(\cos\phi)^2]=0 .
\end{equation}
To make it clear, we introduce $x=\cos^2 \theta $, $1-x=\sin^2\theta$, $y=\cos^2\phi$, $1-y=\sin^2\phi$ and obtain 
\begin{equation}\label{c_n=0}
    d\tilde e_0+\sum_{n=0}^\infty \tilde d_nF_n(x,y)=0 ,
\end{equation}
where $F_n(x,y)=[(1-x)^{n+1}y^n(1-y)+x^{n+1}y^n(1-y)+y(1-y)^n]$, 
and the above equation holds for any $0<x,y<1$. 
For a specific value $x=\frac 12$, we have 
\begin{equation}
    F_n(\frac 12,y)=\frac{1}{2^n}y^n(1-y)+y(1-y)^n .
\end{equation}
For $n>0$, $F_n(1/2,y)$ is a polynomial of $y$ with degree $n+1$, while for $n=0$, $F_0=1$. 
Therefore, $F_n$ with different $n$ are linearly independent. 
According to Eq. \eqref{c_n=0}, we find $\tilde d_n=0$ for $n>0$, and $\tilde d_0=-d\tilde e_0$. 
Here, $d$ is the dimension of the Hilbert space. 
If considering $\sum_{j=1}^d \ev{\tilde N_{1j}}{\phi_{2}}=0$, the similar derivation leads to $\tilde c_n=0,n>0$ and $\tilde c_0=-d\tilde e_0$. 

Next, we investigate the following matrix element to find the relation of $\tilde a_n$, 
\begin{equation}
    \sum_{j=1}^d \mel{\psi_2}{\tilde N_{j1}}{\psi_1}=0 .
\end{equation}
In explicit form, we have 
\begin{equation}
    \sum_{n=1}^\infty \tilde a_n[(\sin\theta\cos\phi)^{2n-1}\sin\theta\sin\phi+(\cos\theta\cos\phi)^{2n-1}\cos\theta\sin\phi+(-\sin\phi)^{2n-1}\cos\phi]=0 .
\end{equation}
As long as $0<\phi<\pi/2$, we can eliminate the common factor $\sin\phi\cos\phi$ in the above equation 
and obtain 
\begin{equation}\label{sum_a}
    \sum_{n=1}^\infty \tilde a_n G_n(x,y) =0, 
\end{equation}
where $G_n(x,y)=[x^ny^{n-1}+(1-x)^ny^{n-1}-(1-y)^{n-1}]$, and $x=\cos^2\theta,y=\cos^2\phi$ again. 
The above equation is valid for any $0<x,y<1$. 
We find that $G_1=0$ for any value of $x,y$. 
Assigning a specific value $x=\frac 12$, $G_n$ reduces to 
\begin{equation}
    G_n(\frac 12,y)=\frac{1}{2^{n-1}} y^{n-1}-(1-y)^{n-1} .
\end{equation}
For $n>1$, $G_n(1/2,y)$ is a polynomial of $y$ with degree $n-1$. 
$\{G_n\}, (n>1)$ are linearly independent, so $\tilde a_n=0$ for $n>1$ according to Eq. \eqref{sum_a}. 
Since $G_1=0$, this equation puts no constraint on $\tilde a_1$. 
Likewise, if we consider $\sum_{j=1}^d \mel{\psi_1}{\tilde N_{j1}}{\psi_2}=0$ instead, 
we can deduce that $\tilde b_n=0$ for $n>1$ while $\tilde b_1$ is free. 

The remaining unknown coefficients are $\tilde a_1,\tilde b_1$. 
Let us consider 
\begin{equation}
    \sum_{j=1}^d \ev{\tilde N_{j1}}{\psi_1}=0 .
\end{equation}
After some simple calculations, we find the relation between $\tilde a_1,\tilde b_1$: 
\begin{equation}
    \tilde a_1+\tilde b_1=d^2\tilde e_0 .
\end{equation}

Up to now, the expansion for $\tilde N_{ji}$ is fixed to
\begin{equation}
    \tilde N_{ji}(\Pi^1_j,\Pi^0_i)=\tilde e_0 I-d\tilde e_0[\Pi^0_i+\Pi^1_j]+\tilde a_1\Pi^1_j\Pi^0_i+(-\tilde a_1+d^2\tilde e_0)\Pi^0_i\Pi^1_j ,
\end{equation}
with free coefficients $\tilde e_0,\tilde a_1$. 
On the other hand, it can be checked that the above expression satisfies the condition for $\tilde N_{ji}$ in Eq. \eqref{ding}, 
so it may be regarded as the general form of $\tilde N_{ji}$. 
However, we hope the definition of work is universal and irrespective of the dimension of Hilbert space $d$. 
Especially, the above expression should work well in infinite-dimensional space. 
Therefore, we are forced to choose $\tilde e_0=0$. 
By the way, we find that 2-dimensional Hilbert space is an exception because we have less freedom to choose the projective operators. 
Nevertheless, since the definition should be universal irrespective of the dimension, this exception in 2d space does not matter. 

In conclusion, under the condition from the first law, the general form for $N_{ji}$ is simply
\begin{equation}
    N_{ji}(\Pi^1_j,\Pi^0_i)=\frac{1+\eta}{2}\Pi^1_j\Pi^0_i+\frac{1-\eta}{2}\Pi^0_i\Pi^1_j ,
\end{equation}
where $\eta=2\tilde a_1$ can be arbitrary complex number. 
The corresponding distribution function reads
\begin{equation}\label{r_class}
    P(w)=\sum_{ji}\delta(w-(\epsilon^1_j-\epsilon^0_i))\Tr \left[ \left(\frac{1+\eta}{2}\Pi^1_j\Pi^0_i+\frac{1-\eta}{2}\Pi^0_i\Pi^1_j \right)\rho \right] .
\end{equation}
This form contains the MH quasiprobability for work ($\eta=0$) and the KD quasiprobability for work ($\eta=1$).

At the end of this appendix, we will show that the above distribution satisfies time-reversal symmetry (E2). 
In the time-reversed process described in requirement (E2), we have the following relations in the Heisenberg picture: 
$\bar H_1=\Theta U H_0 U^{-1}\Theta^{-1}, \bar H_0=\Theta U H_1 U^{-1}\Theta^{-1}$, $\bar\rho=\Theta U\rho U^{-1}\Theta^{-1}$. 
The eigenvalues and corresponding projective operators are: $\bar\epsilon^0_i=\epsilon^1_i,\bar\epsilon^1_j=\epsilon^0_j$, 
$\bar\Pi^0_i=\Theta U\Pi^1_i U^{-1}\Theta^{-1} ,\bar\Pi^1_j=\Theta U\Pi^0_j U^{-1}\Theta^{-1}$. 
Operator $N_{ji}$ now changes to 
$$ N_{ji}(\bar\Pi^1_j,\bar\Pi^0_i)=\Theta U N_{ji}^*(\Pi^0_j,\Pi^1_i) U^{-1}\Theta^{-1},$$ 
where superscript $*$ on $N_{ji}$ denotes the complex conjugate over all the coefficients in the expansion of $N_{ji}$. 
After some derivation, we get the distribution function in the time-reversed process: 
\begin{equation}
    \begin{aligned}
    \bar P(w)&=\sum_{ji}\delta(w-(\bar\epsilon^1_j-\bar\epsilon^0_i)) \Tr[N_{ji}(\bar\Pi^1_j,\bar\Pi^0_i)\bar\rho] \\
    &=\sum_{ji}\delta(w-(\epsilon^0_j-\epsilon^1_i)) \Tr[N_{ji}^*(\Pi^0_j,\Pi^1_i) \rho]^* \\
    &=\sum_{ji}\delta(w+(\epsilon^1_j-\epsilon^0_i)) \Tr[N_{ij}^{*\dagger}(\Pi^0_i,\Pi^1_j) \rho] .
    \end{aligned}
\end{equation}
In the second equality, please notice that when canceling the anti-unitary operator $\Theta$ in the trace, there will be an extra complex conjugate. 
In the third equality, we exchange the summation index $i,j$. 
Imposing $\bar P(w)= P(-w)$, we find that the time-reversal condition on $N_{ji}$ is 
\begin{equation}
    N_{ij}^{*\dagger}(\Pi^0_i,\Pi^1_j)=N_{ji}(\Pi^1_j,\Pi^0_i) .
\end{equation}
In terms of expansion coefficients in Eq. \eqref{expansion_1}, we find that $c_n=d_n$ while there are no constraints on $a_n,b_n$. 
$N_{ji}$ in Eq. \eqref{r_class} satisfies this condition. 
Particularly, both the MH quasiprobability of work and the KD quasiprobability of work encode the time-reversal symmetry. 
Thus, the time-reversal symmetry property (E2) can be regarded as a consequence of other requirements. 

\end{widetext}
\bibliography{choice}

\begin{thebibliography}{65}%
\makeatletter
\providecommand \@ifxundefined [1]{%
 \@ifx{#1\undefined}
}%
\providecommand \@ifnum [1]{%
 \ifnum #1\expandafter \@firstoftwo
 \else \expandafter \@secondoftwo
 \fi
}%
\providecommand \@ifx [1]{%
 \ifx #1\expandafter \@firstoftwo
 \else \expandafter \@secondoftwo
 \fi
}%
\providecommand \natexlab [1]{#1}%
\providecommand \enquote  [1]{``#1''}%
\providecommand \bibnamefont  [1]{#1}%
\providecommand \bibfnamefont [1]{#1}%
\providecommand \citenamefont [1]{#1}%
\providecommand \href@noop [0]{\@secondoftwo}%
\providecommand \href [0]{\begingroup \@sanitize@url \@href}%
\providecommand \@href[1]{\@@startlink{#1}\@@href}%
\providecommand \@@href[1]{\endgroup#1\@@endlink}%
\providecommand \@sanitize@url [0]{\catcode `\\12\catcode `\$12\catcode
  `\&12\catcode `\#12\catcode `\^12\catcode `\_12\catcode `\%12\relax}%
\providecommand \@@startlink[1]{}%
\providecommand \@@endlink[0]{}%
\providecommand \url  [0]{\begingroup\@sanitize@url \@url }%
\providecommand \@url [1]{\endgroup\@href {#1}{\urlprefix }}%
\providecommand \urlprefix  [0]{URL }%
\providecommand \Eprint [0]{\href }%
\providecommand \doibase [0]{https://doi.org/}%
\providecommand \selectlanguage [0]{\@gobble}%
\providecommand \bibinfo  [0]{\@secondoftwo}%
\providecommand \bibfield  [0]{\@secondoftwo}%
\providecommand \translation [1]{[#1]}%
\providecommand \BibitemOpen [0]{}%
\providecommand \bibitemStop [0]{}%
\providecommand \bibitemNoStop [0]{.\EOS\space}%
\providecommand \EOS [0]{\spacefactor3000\relax}%
\providecommand \BibitemShut  [1]{\csname bibitem#1\endcsname}%
\let\auto@bib@innerbib\@empty
\bibitem [{\citenamefont {Seifert}(2012)}]{Seifert_2012}%
  \BibitemOpen
  \bibfield  {author} {\bibinfo {author} {\bibfnamefont {U.}~\bibnamefont
  {Seifert}},\ }\bibfield  {title} {\bibinfo {title} {Stochastic
  thermodynamics, fluctuation theorems and molecular machines},\ }\href
  {https://doi.org/10.1088/0034-4885/75/12/126001} {\bibfield  {journal}
  {\bibinfo  {journal} {Rep. Prog. Phys.}\ }\textbf {\bibinfo {volume} {75}},\
  \bibinfo {pages} {126001} (\bibinfo {year} {2012})}\BibitemShut {NoStop}%
\bibitem [{\citenamefont {{Peliti}}\ and\ \citenamefont
  {{Pigolotti}}(2021)}]{book_ST}%
  \BibitemOpen
  \bibfield  {author} {\bibinfo {author} {\bibfnamefont {L.}~\bibnamefont
  {{Peliti}}}\ and\ \bibinfo {author} {\bibfnamefont {S.}~\bibnamefont
  {{Pigolotti}}},\ }\href@noop {} {\emph {\bibinfo {title} {{Stochastic
  Thermodynamics: An Introduction}}}}\ (\bibinfo  {publisher} {Princeton
  University Press},\ \bibinfo {year} {2021})\BibitemShut {NoStop}%
\bibitem [{\citenamefont {Sekimoto}(2010)}]{sekimoto2010stochastic}%
  \BibitemOpen
  \bibfield  {author} {\bibinfo {author} {\bibfnamefont {K.}~\bibnamefont
  {Sekimoto}},\ }\href@noop {} {\emph {\bibinfo {title} {Stochastic
  energetics}}},\ Vol.\ \bibinfo {volume} {799}\ (\bibinfo  {publisher}
  {Springer},\ \bibinfo {year} {2010})\BibitemShut {NoStop}%
\bibitem [{\citenamefont {Jarzynski}(2011)}]{FTs}%
  \BibitemOpen
  \bibfield  {author} {\bibinfo {author} {\bibfnamefont {C.}~\bibnamefont
  {Jarzynski}},\ }\bibfield  {title} {\bibinfo {title} {Equalities and
  inequalities: Irreversibility and the second law of thermodynamics at the
  nanoscale},\ }\href
  {https://doi.org/10.1146/annurev-conmatphys-062910-140506} {\bibfield
  {journal} {\bibinfo  {journal} {Annu. Rev. Conden. Matt. Phys.}\ }\textbf
  {\bibinfo {volume} {2}},\ \bibinfo {pages} {329} (\bibinfo {year}
  {2011})}\BibitemShut {NoStop}%
\bibitem [{\citenamefont {Vinjanampathy}\ and\ \citenamefont
  {Anders}(2016)}]{Sai2016}%
  \BibitemOpen
  \bibfield  {author} {\bibinfo {author} {\bibfnamefont {S.}~\bibnamefont
  {Vinjanampathy}}\ and\ \bibinfo {author} {\bibfnamefont {J.}~\bibnamefont
  {Anders}},\ }\bibfield  {title} {\bibinfo {title} {Quantum thermodynamics},\
  }\href {https://doi.org/10.1080/00107514.2016.1201896} {\bibfield  {journal}
  {\bibinfo  {journal} {Contemp. Phys.}\ }\textbf {\bibinfo {volume} {57}},\
  \bibinfo {pages} {545} (\bibinfo {year} {2016})},\ \Eprint
  {https://arxiv.org/abs/https://doi.org/10.1080/00107514.2016.1201896}
  {https://doi.org/10.1080/00107514.2016.1201896} \BibitemShut {NoStop}%
\bibitem [{\citenamefont {Pekola}(2015)}]{Pekola2015}%
  \BibitemOpen
  \bibfield  {author} {\bibinfo {author} {\bibfnamefont {J.~P.}\ \bibnamefont
  {Pekola}},\ }\bibfield  {title} {\bibinfo {title} {Towards quantum
  thermodynamics in electronic circuits},\ }\href
  {https://doi.org/10.1038/nphys3169} {\bibfield  {journal} {\bibinfo
  {journal} {Nat. Phys.}\ }\textbf {\bibinfo {volume} {11}},\ \bibinfo {pages}
  {118} (\bibinfo {year} {2015})}\BibitemShut {NoStop}%
\bibitem [{\citenamefont {Strasberg}(2022)}]{book_ox}%
  \BibitemOpen
  \bibfield  {author} {\bibinfo {author} {\bibfnamefont {P.}~\bibnamefont
  {Strasberg}},\ }\href {https://doi.org/10.1093/oso/9780192895585.001.0001}
  {\emph {\bibinfo {title} {{Quantum Stochastic Thermodynamics: Foundations and
  Selected Applications}}}}\ (\bibinfo  {publisher} {Oxford University Press},\
  \bibinfo {year} {2022})\ \Eprint
  {https://arxiv.org/abs/https://academic.oup.com/book/43799/book-pdf/50145705/9780192648143\_web.pdf}
  {https://academic.oup.com/book/43799/book-pdf/50145705/9780192648143\_web.pdf}
  \BibitemShut {NoStop}%
\bibitem [{\citenamefont {Goold}\ \emph {et~al.}(2016)\citenamefont {Goold},
  \citenamefont {Huber}, \citenamefont {Riera}, \citenamefont {del Rio},\ and\
  \citenamefont {Skrzypczyk}}]{Goold_2016}%
  \BibitemOpen
  \bibfield  {author} {\bibinfo {author} {\bibfnamefont {J.}~\bibnamefont
  {Goold}}, \bibinfo {author} {\bibfnamefont {M.}~\bibnamefont {Huber}},
  \bibinfo {author} {\bibfnamefont {A.}~\bibnamefont {Riera}}, \bibinfo
  {author} {\bibfnamefont {L.}~\bibnamefont {del Rio}},\ and\ \bibinfo {author}
  {\bibfnamefont {P.}~\bibnamefont {Skrzypczyk}},\ }\bibfield  {title}
  {\bibinfo {title} {The role of quantum information in thermodynamics—a
  topical review},\ }\href {https://doi.org/10.1088/1751-8113/49/14/143001}
  {\bibfield  {journal} {\bibinfo  {journal} {J. Phys. A: Math. Theor.}\
  }\textbf {\bibinfo {volume} {49}},\ \bibinfo {pages} {143001} (\bibinfo
  {year} {2016})}\BibitemShut {NoStop}%
\bibitem [{\citenamefont {\AA{}berg}(2018)}]{PhysRevX.8.011019}%
  \BibitemOpen
  \bibfield  {author} {\bibinfo {author} {\bibfnamefont {J.}~\bibnamefont
  {\AA{}berg}},\ }\bibfield  {title} {\bibinfo {title} {Fully quantum
  fluctuation theorems},\ }\href {https://doi.org/10.1103/PhysRevX.8.011019}
  {\bibfield  {journal} {\bibinfo  {journal} {Phys. Rev. X}\ }\textbf {\bibinfo
  {volume} {8}},\ \bibinfo {pages} {011019} (\bibinfo {year}
  {2018})}\BibitemShut {NoStop}%
\bibitem [{\citenamefont {Campisi}\ \emph {et~al.}(2015)\citenamefont
  {Campisi}, \citenamefont {Pekola},\ and\ \citenamefont
  {Fazio}}]{Campisi_2015}%
  \BibitemOpen
  \bibfield  {author} {\bibinfo {author} {\bibfnamefont {M.}~\bibnamefont
  {Campisi}}, \bibinfo {author} {\bibfnamefont {J.}~\bibnamefont {Pekola}},\
  and\ \bibinfo {author} {\bibfnamefont {R.}~\bibnamefont {Fazio}},\ }\bibfield
   {title} {\bibinfo {title} {Nonequilibrium fluctuations in quantum heat
  engines: theory, example, and possible solid state experiments},\ }\href
  {https://doi.org/10.1088/1367-2630/17/3/035012} {\bibfield  {journal}
  {\bibinfo  {journal} {New J. Phys.}\ }\textbf {\bibinfo {volume} {17}},\
  \bibinfo {pages} {035012} (\bibinfo {year} {2015})}\BibitemShut {NoStop}%
\bibitem [{\citenamefont {Millen}\ and\ \citenamefont
  {Xuereb}(2016)}]{Millen_2016}%
  \BibitemOpen
  \bibfield  {author} {\bibinfo {author} {\bibfnamefont {J.}~\bibnamefont
  {Millen}}\ and\ \bibinfo {author} {\bibfnamefont {A.}~\bibnamefont
  {Xuereb}},\ }\bibfield  {title} {\bibinfo {title} {Perspective on quantum
  thermodynamics},\ }\href {https://doi.org/10.1088/1367-2630/18/1/011002}
  {\bibfield  {journal} {\bibinfo  {journal} {New J. Phys.}\ }\textbf {\bibinfo
  {volume} {18}},\ \bibinfo {pages} {011002} (\bibinfo {year}
  {2016})}\BibitemShut {NoStop}%
\bibitem [{\citenamefont {Manzano}\ \emph {et~al.}(2018)\citenamefont
  {Manzano}, \citenamefont {Horowitz},\ and\ \citenamefont
  {Parrondo}}]{PhysRevX.8.031037}%
  \BibitemOpen
  \bibfield  {author} {\bibinfo {author} {\bibfnamefont {G.}~\bibnamefont
  {Manzano}}, \bibinfo {author} {\bibfnamefont {J.~M.}\ \bibnamefont
  {Horowitz}},\ and\ \bibinfo {author} {\bibfnamefont {J.~M.~R.}\ \bibnamefont
  {Parrondo}},\ }\bibfield  {title} {\bibinfo {title} {Quantum fluctuation
  theorems for arbitrary environments: Adiabatic and nonadiabatic entropy
  production},\ }\href {https://doi.org/10.1103/PhysRevX.8.031037} {\bibfield
  {journal} {\bibinfo  {journal} {Phys. Rev. X}\ }\textbf {\bibinfo {volume}
  {8}},\ \bibinfo {pages} {031037} (\bibinfo {year} {2018})}\BibitemShut
  {NoStop}%
\bibitem [{\citenamefont {Talkner}\ and\ \citenamefont
  {H\"anggi}(2016)}]{PhysRevE.93.022131}%
  \BibitemOpen
  \bibfield  {author} {\bibinfo {author} {\bibfnamefont {P.}~\bibnamefont
  {Talkner}}\ and\ \bibinfo {author} {\bibfnamefont {P.}~\bibnamefont
  {H\"anggi}},\ }\bibfield  {title} {\bibinfo {title} {Aspects of quantum
  work},\ }\href {https://doi.org/10.1103/PhysRevE.93.022131} {\bibfield
  {journal} {\bibinfo  {journal} {Phys. Rev. E}\ }\textbf {\bibinfo {volume}
  {93}},\ \bibinfo {pages} {022131} (\bibinfo {year} {2016})}\BibitemShut
  {NoStop}%
\bibitem [{\citenamefont {Bochkov}\ and\ \citenamefont
  {Kuzovlev}(1977)}]{o-of-work_77}%
  \BibitemOpen
  \bibfield  {author} {\bibinfo {author} {\bibfnamefont {G.}~\bibnamefont
  {Bochkov}}\ and\ \bibinfo {author} {\bibfnamefont {Y.}~\bibnamefont
  {Kuzovlev}},\ }\bibfield  {title} {\bibinfo {title} {General theory of
  thermal fluctuations in nonlinear systems},\ }\href@noop {} {\bibfield
  {journal} {\bibinfo  {journal} {Zh. Eksp. Teor. Fiz}\ }\textbf {\bibinfo
  {volume} {72}},\ \bibinfo {pages} {238} (\bibinfo {year} {1977})}\BibitemShut
  {NoStop}%
\bibitem [{\citenamefont {Allahverdyan}\ and\ \citenamefont
  {Nieuwenhuizen}(2005)}]{o-of-work_05}%
  \BibitemOpen
  \bibfield  {author} {\bibinfo {author} {\bibfnamefont {A.~E.}\ \bibnamefont
  {Allahverdyan}}\ and\ \bibinfo {author} {\bibfnamefont {T.~M.}\ \bibnamefont
  {Nieuwenhuizen}},\ }\bibfield  {title} {\bibinfo {title} {Fluctuations of
  work from quantum subensembles: The case against quantum work-fluctuation
  theorems},\ }\href {https://doi.org/10.1103/PhysRevE.71.066102} {\bibfield
  {journal} {\bibinfo  {journal} {Phys. Rev. E}\ }\textbf {\bibinfo {volume}
  {71}},\ \bibinfo {pages} {066102} (\bibinfo {year} {2005})}\BibitemShut
  {NoStop}%
\bibitem [{\citenamefont {Kurchan}(2001)}]{TPM1}%
  \BibitemOpen
  \bibfield  {author} {\bibinfo {author} {\bibfnamefont {J.}~\bibnamefont
  {Kurchan}},\ }\href@noop {} {\bibinfo {title} {A quantum fluctuation
  theorem}} (\bibinfo {year} {2001}),\ \Eprint
  {https://arxiv.org/abs/cond-mat/0007360} {arXiv:cond-mat/0007360
  [cond-mat.stat-mech]} \BibitemShut {NoStop}%
\bibitem [{\citenamefont {Tasaki}(2000)}]{TPM2}%
  \BibitemOpen
  \bibfield  {author} {\bibinfo {author} {\bibfnamefont {H.}~\bibnamefont
  {Tasaki}},\ }\href@noop {} {\bibinfo {title} {Jarzynski relations for quantum
  systems and some applications}} (\bibinfo {year} {2000}),\ \Eprint
  {https://arxiv.org/abs/cond-mat/0009244} {arXiv:cond-mat/0009244
  [cond-mat.stat-mech]} \BibitemShut {NoStop}%
\bibitem [{\citenamefont {Talkner}\ \emph {et~al.}(2009)\citenamefont
  {Talkner}, \citenamefont {Campisi},\ and\ \citenamefont
  {Hänggi}}]{Talkner_2009}%
  \BibitemOpen
  \bibfield  {author} {\bibinfo {author} {\bibfnamefont {P.}~\bibnamefont
  {Talkner}}, \bibinfo {author} {\bibfnamefont {M.}~\bibnamefont {Campisi}},\
  and\ \bibinfo {author} {\bibfnamefont {P.}~\bibnamefont {Hänggi}},\
  }\bibfield  {title} {\bibinfo {title} {Fluctuation theorems in driven open
  quantum systems},\ }\href {https://doi.org/10.1088/1742-5468/2009/02/P02025}
  {\bibfield  {journal} {\bibinfo  {journal} {J. Stat. Mech: Theory Exp.}\
  }\textbf {\bibinfo {volume} {2009}},\ \bibinfo {pages} {P02025} (\bibinfo
  {year} {2009})}\BibitemShut {NoStop}%
\bibitem [{\citenamefont {Esposito}\ \emph {et~al.}(2009)\citenamefont
  {Esposito}, \citenamefont {Harbola},\ and\ \citenamefont
  {Mukamel}}]{RevModPhys.81.1665}%
  \BibitemOpen
  \bibfield  {author} {\bibinfo {author} {\bibfnamefont {M.}~\bibnamefont
  {Esposito}}, \bibinfo {author} {\bibfnamefont {U.}~\bibnamefont {Harbola}},\
  and\ \bibinfo {author} {\bibfnamefont {S.}~\bibnamefont {Mukamel}},\
  }\bibfield  {title} {\bibinfo {title} {Nonequilibrium fluctuations,
  fluctuation theorems, and counting statistics in quantum systems},\ }\href
  {https://doi.org/10.1103/RevModPhys.81.1665} {\bibfield  {journal} {\bibinfo
  {journal} {Rev. Mod. Phys.}\ }\textbf {\bibinfo {volume} {81}},\ \bibinfo
  {pages} {1665} (\bibinfo {year} {2009})}\BibitemShut {NoStop}%
\bibitem [{\citenamefont {Campisi}\ \emph {et~al.}(2011)\citenamefont
  {Campisi}, \citenamefont {H\"anggi},\ and\ \citenamefont
  {Talkner}}]{RevModPhys.83.771}%
  \BibitemOpen
  \bibfield  {author} {\bibinfo {author} {\bibfnamefont {M.}~\bibnamefont
  {Campisi}}, \bibinfo {author} {\bibfnamefont {P.}~\bibnamefont {H\"anggi}},\
  and\ \bibinfo {author} {\bibfnamefont {P.}~\bibnamefont {Talkner}},\
  }\bibfield  {title} {\bibinfo {title} {Colloquium: Quantum fluctuation
  relations: Foundations and applications},\ }\href
  {https://doi.org/10.1103/RevModPhys.83.771} {\bibfield  {journal} {\bibinfo
  {journal} {Rev. Mod. Phys.}\ }\textbf {\bibinfo {volume} {83}},\ \bibinfo
  {pages} {771} (\bibinfo {year} {2011})}\BibitemShut {NoStop}%
\bibitem [{\citenamefont {Batalh\~ao}\ \emph {et~al.}(2014)\citenamefont
  {Batalh\~ao}, \citenamefont {Souza}, \citenamefont {Mazzola}, \citenamefont
  {Auccaise}, \citenamefont {Sarthour}, \citenamefont {Oliveira}, \citenamefont
  {Goold}, \citenamefont {De~Chiara}, \citenamefont {Paternostro},\ and\
  \citenamefont {Serra}}]{exp_work}%
  \BibitemOpen
  \bibfield  {author} {\bibinfo {author} {\bibfnamefont {T.~B.}\ \bibnamefont
  {Batalh\~ao}}, \bibinfo {author} {\bibfnamefont {A.~M.}\ \bibnamefont
  {Souza}}, \bibinfo {author} {\bibfnamefont {L.}~\bibnamefont {Mazzola}},
  \bibinfo {author} {\bibfnamefont {R.}~\bibnamefont {Auccaise}}, \bibinfo
  {author} {\bibfnamefont {R.~S.}\ \bibnamefont {Sarthour}}, \bibinfo {author}
  {\bibfnamefont {I.~S.}\ \bibnamefont {Oliveira}}, \bibinfo {author}
  {\bibfnamefont {J.}~\bibnamefont {Goold}}, \bibinfo {author} {\bibfnamefont
  {G.}~\bibnamefont {De~Chiara}}, \bibinfo {author} {\bibfnamefont
  {M.}~\bibnamefont {Paternostro}},\ and\ \bibinfo {author} {\bibfnamefont
  {R.~M.}\ \bibnamefont {Serra}},\ }\bibfield  {title} {\bibinfo {title}
  {Experimental reconstruction of work distribution and study of fluctuation
  relations in a closed quantum system},\ }\href
  {https://doi.org/10.1103/PhysRevLett.113.140601} {\bibfield  {journal}
  {\bibinfo  {journal} {Phys. Rev. Lett.}\ }\textbf {\bibinfo {volume} {113}},\
  \bibinfo {pages} {140601} (\bibinfo {year} {2014})}\BibitemShut {NoStop}%
\bibitem [{\citenamefont {An}\ \emph {et~al.}(2015)\citenamefont {An},
  \citenamefont {Zhang}, \citenamefont {Um}, \citenamefont {Lv}, \citenamefont
  {Lu}, \citenamefont {Zhang}, \citenamefont {Yin}, \citenamefont {Quan},\ and\
  \citenamefont {Kim}}]{An2015}%
  \BibitemOpen
  \bibfield  {author} {\bibinfo {author} {\bibfnamefont {S.}~\bibnamefont
  {An}}, \bibinfo {author} {\bibfnamefont {J.-N.}\ \bibnamefont {Zhang}},
  \bibinfo {author} {\bibfnamefont {M.}~\bibnamefont {Um}}, \bibinfo {author}
  {\bibfnamefont {D.}~\bibnamefont {Lv}}, \bibinfo {author} {\bibfnamefont
  {Y.}~\bibnamefont {Lu}}, \bibinfo {author} {\bibfnamefont {J.}~\bibnamefont
  {Zhang}}, \bibinfo {author} {\bibfnamefont {Z.-Q.}\ \bibnamefont {Yin}},
  \bibinfo {author} {\bibfnamefont {H.~T.}\ \bibnamefont {Quan}},\ and\
  \bibinfo {author} {\bibfnamefont {K.}~\bibnamefont {Kim}},\ }\bibfield
  {title} {\bibinfo {title} {Experimental test of the quantum jarzynski
  equality with a trapped-ion system},\ }\href
  {https://doi.org/10.1038/nphys3197} {\bibfield  {journal} {\bibinfo
  {journal} {Nat. Phys.}\ }\textbf {\bibinfo {volume} {11}},\ \bibinfo {pages}
  {193} (\bibinfo {year} {2015})}\BibitemShut {NoStop}%
\bibitem [{\citenamefont {Ro\ss{}nagel}\ \emph {et~al.}(2014)\citenamefont
  {Ro\ss{}nagel}, \citenamefont {Abah}, \citenamefont {Schmidt-Kaler},
  \citenamefont {Singer},\ and\ \citenamefont {Lutz}}]{QHE_Carnot}%
  \BibitemOpen
  \bibfield  {author} {\bibinfo {author} {\bibfnamefont {J.}~\bibnamefont
  {Ro\ss{}nagel}}, \bibinfo {author} {\bibfnamefont {O.}~\bibnamefont {Abah}},
  \bibinfo {author} {\bibfnamefont {F.}~\bibnamefont {Schmidt-Kaler}}, \bibinfo
  {author} {\bibfnamefont {K.}~\bibnamefont {Singer}},\ and\ \bibinfo {author}
  {\bibfnamefont {E.}~\bibnamefont {Lutz}},\ }\bibfield  {title} {\bibinfo
  {title} {Nanoscale heat engine beyond the carnot limit},\ }\href
  {https://doi.org/10.1103/PhysRevLett.112.030602} {\bibfield  {journal}
  {\bibinfo  {journal} {Phys. Rev. Lett.}\ }\textbf {\bibinfo {volume} {112}},\
  \bibinfo {pages} {030602} (\bibinfo {year} {2014})}\BibitemShut {NoStop}%
\bibitem [{\citenamefont {Brandner}\ \emph {et~al.}(2017)\citenamefont
  {Brandner}, \citenamefont {Bauer},\ and\ \citenamefont
  {Seifert}}]{QHE_linear}%
  \BibitemOpen
  \bibfield  {author} {\bibinfo {author} {\bibfnamefont {K.}~\bibnamefont
  {Brandner}}, \bibinfo {author} {\bibfnamefont {M.}~\bibnamefont {Bauer}},\
  and\ \bibinfo {author} {\bibfnamefont {U.}~\bibnamefont {Seifert}},\
  }\bibfield  {title} {\bibinfo {title} {Universal coherence-induced power
  losses of quantum heat engines in linear response},\ }\href
  {https://doi.org/10.1103/PhysRevLett.119.170602} {\bibfield  {journal}
  {\bibinfo  {journal} {Phys. Rev. Lett.}\ }\textbf {\bibinfo {volume} {119}},\
  \bibinfo {pages} {170602} (\bibinfo {year} {2017})}\BibitemShut {NoStop}%
\bibitem [{\citenamefont {Klatzow}\ \emph {et~al.}(2019)\citenamefont
  {Klatzow}, \citenamefont {Becker}, \citenamefont {Ledingham}, \citenamefont
  {Weinzetl}, \citenamefont {Kaczmarek}, \citenamefont {Saunders},
  \citenamefont {Nunn}, \citenamefont {Walmsley}, \citenamefont {Uzdin},\ and\
  \citenamefont {Poem}}]{QHE_experiment}%
  \BibitemOpen
  \bibfield  {author} {\bibinfo {author} {\bibfnamefont {J.}~\bibnamefont
  {Klatzow}}, \bibinfo {author} {\bibfnamefont {J.~N.}\ \bibnamefont {Becker}},
  \bibinfo {author} {\bibfnamefont {P.~M.}\ \bibnamefont {Ledingham}}, \bibinfo
  {author} {\bibfnamefont {C.}~\bibnamefont {Weinzetl}}, \bibinfo {author}
  {\bibfnamefont {K.~T.}\ \bibnamefont {Kaczmarek}}, \bibinfo {author}
  {\bibfnamefont {D.~J.}\ \bibnamefont {Saunders}}, \bibinfo {author}
  {\bibfnamefont {J.}~\bibnamefont {Nunn}}, \bibinfo {author} {\bibfnamefont
  {I.~A.}\ \bibnamefont {Walmsley}}, \bibinfo {author} {\bibfnamefont
  {R.}~\bibnamefont {Uzdin}},\ and\ \bibinfo {author} {\bibfnamefont
  {E.}~\bibnamefont {Poem}},\ }\bibfield  {title} {\bibinfo {title}
  {Experimental demonstration of quantum effects in the operation of
  microscopic heat engines},\ }\href
  {https://doi.org/10.1103/PhysRevLett.122.110601} {\bibfield  {journal}
  {\bibinfo  {journal} {Phys. Rev. Lett.}\ }\textbf {\bibinfo {volume} {122}},\
  \bibinfo {pages} {110601} (\bibinfo {year} {2019})}\BibitemShut {NoStop}%
\bibitem [{\citenamefont {Uzdin}\ \emph {et~al.}(2015)\citenamefont {Uzdin},
  \citenamefont {Levy},\ and\ \citenamefont {Kosloff}}]{QHE_extract}%
  \BibitemOpen
  \bibfield  {author} {\bibinfo {author} {\bibfnamefont {R.}~\bibnamefont
  {Uzdin}}, \bibinfo {author} {\bibfnamefont {A.}~\bibnamefont {Levy}},\ and\
  \bibinfo {author} {\bibfnamefont {R.}~\bibnamefont {Kosloff}},\ }\bibfield
  {title} {\bibinfo {title} {Equivalence of quantum heat machines, and
  quantum-thermodynamic signatures},\ }\href
  {https://doi.org/10.1103/PhysRevX.5.031044} {\bibfield  {journal} {\bibinfo
  {journal} {Phys. Rev. X}\ }\textbf {\bibinfo {volume} {5}},\ \bibinfo {pages}
  {031044} (\bibinfo {year} {2015})}\BibitemShut {NoStop}%
\bibitem [{\citenamefont {Watanabe}\ \emph {et~al.}(2017)\citenamefont
  {Watanabe}, \citenamefont {Venkatesh}, \citenamefont {Talkner},\ and\
  \citenamefont {del Campo}}]{many_cycles}%
  \BibitemOpen
  \bibfield  {author} {\bibinfo {author} {\bibfnamefont {G.}~\bibnamefont
  {Watanabe}}, \bibinfo {author} {\bibfnamefont {B.~P.}\ \bibnamefont
  {Venkatesh}}, \bibinfo {author} {\bibfnamefont {P.}~\bibnamefont {Talkner}},\
  and\ \bibinfo {author} {\bibfnamefont {A.}~\bibnamefont {del Campo}},\
  }\bibfield  {title} {\bibinfo {title} {Quantum performance of thermal
  machines over many cycles},\ }\href
  {https://doi.org/10.1103/PhysRevLett.118.050601} {\bibfield  {journal}
  {\bibinfo  {journal} {Phys. Rev. Lett.}\ }\textbf {\bibinfo {volume} {118}},\
  \bibinfo {pages} {050601} (\bibinfo {year} {2017})}\BibitemShut {NoStop}%
\bibitem [{\citenamefont {Ma}\ \emph {et~al.}(2021)\citenamefont {Ma},
  \citenamefont {Liu},\ and\ \citenamefont {Sun}}]{MYH}%
  \BibitemOpen
  \bibfield  {author} {\bibinfo {author} {\bibfnamefont {Y.-H.}\ \bibnamefont
  {Ma}}, \bibinfo {author} {\bibfnamefont {C.~L.}\ \bibnamefont {Liu}},\ and\
  \bibinfo {author} {\bibfnamefont {C.~P.}\ \bibnamefont {Sun}},\ }\href@noop
  {} {\bibinfo {title} {Works with quantum resource of coherence}} (\bibinfo
  {year} {2021}),\ \Eprint {https://arxiv.org/abs/2110.04550} {arXiv:2110.04550
  [quant-ph]} \BibitemShut {NoStop}%
\bibitem [{\citenamefont {Chen}\ \emph
  {et~al.}(2019{\natexlab{a}})\citenamefont {Chen}, \citenamefont {Sun},\ and\
  \citenamefont {Dong}}]{otto_1}%
  \BibitemOpen
  \bibfield  {author} {\bibinfo {author} {\bibfnamefont {J.-F.}\ \bibnamefont
  {Chen}}, \bibinfo {author} {\bibfnamefont {C.-P.}\ \bibnamefont {Sun}},\ and\
  \bibinfo {author} {\bibfnamefont {H.}~\bibnamefont {Dong}},\ }\bibfield
  {title} {\bibinfo {title} {Boosting the performance of quantum otto heat
  engines},\ }\href {https://doi.org/10.1103/PhysRevE.100.032144} {\bibfield
  {journal} {\bibinfo  {journal} {Phys. Rev. E}\ }\textbf {\bibinfo {volume}
  {100}},\ \bibinfo {pages} {032144} (\bibinfo {year}
  {2019}{\natexlab{a}})}\BibitemShut {NoStop}%
\bibitem [{\citenamefont {Chen}\ \emph
  {et~al.}(2019{\natexlab{b}})\citenamefont {Chen}, \citenamefont {Sun},\ and\
  \citenamefont {Dong}}]{otto_2}%
  \BibitemOpen
  \bibfield  {author} {\bibinfo {author} {\bibfnamefont {J.-F.}\ \bibnamefont
  {Chen}}, \bibinfo {author} {\bibfnamefont {C.-P.}\ \bibnamefont {Sun}},\ and\
  \bibinfo {author} {\bibfnamefont {H.}~\bibnamefont {Dong}},\ }\bibfield
  {title} {\bibinfo {title} {Achieve higher efficiency at maximum power with
  finite-time quantum otto cycle},\ }\href
  {https://doi.org/10.1103/PhysRevE.100.062140} {\bibfield  {journal} {\bibinfo
   {journal} {Phys. Rev. E}\ }\textbf {\bibinfo {volume} {100}},\ \bibinfo
  {pages} {062140} (\bibinfo {year} {2019}{\natexlab{b}})}\BibitemShut
  {NoStop}%
\bibitem [{\citenamefont {Perarnau-Llobet}\ \emph {et~al.}(2015)\citenamefont
  {Perarnau-Llobet}, \citenamefont {Hovhannisyan}, \citenamefont {Huber},
  \citenamefont {Skrzypczyk}, \citenamefont {Brunner},\ and\ \citenamefont
  {Ac\'{\i}n}}]{extract_correlation}%
  \BibitemOpen
  \bibfield  {author} {\bibinfo {author} {\bibfnamefont {M.}~\bibnamefont
  {Perarnau-Llobet}}, \bibinfo {author} {\bibfnamefont {K.~V.}\ \bibnamefont
  {Hovhannisyan}}, \bibinfo {author} {\bibfnamefont {M.}~\bibnamefont {Huber}},
  \bibinfo {author} {\bibfnamefont {P.}~\bibnamefont {Skrzypczyk}}, \bibinfo
  {author} {\bibfnamefont {N.}~\bibnamefont {Brunner}},\ and\ \bibinfo {author}
  {\bibfnamefont {A.}~\bibnamefont {Ac\'{\i}n}},\ }\bibfield  {title} {\bibinfo
  {title} {Extractable work from correlations},\ }\href
  {https://doi.org/10.1103/PhysRevX.5.041011} {\bibfield  {journal} {\bibinfo
  {journal} {Phys. Rev. X}\ }\textbf {\bibinfo {volume} {5}},\ \bibinfo {pages}
  {041011} (\bibinfo {year} {2015})}\BibitemShut {NoStop}%
\bibitem [{\citenamefont {Korzekwa}\ \emph {et~al.}(2016)\citenamefont
  {Korzekwa}, \citenamefont {Lostaglio}, \citenamefont {Oppenheim},\ and\
  \citenamefont {Jennings}}]{extract_coherence}%
  \BibitemOpen
  \bibfield  {author} {\bibinfo {author} {\bibfnamefont {K.}~\bibnamefont
  {Korzekwa}}, \bibinfo {author} {\bibfnamefont {M.}~\bibnamefont {Lostaglio}},
  \bibinfo {author} {\bibfnamefont {J.}~\bibnamefont {Oppenheim}},\ and\
  \bibinfo {author} {\bibfnamefont {D.}~\bibnamefont {Jennings}},\ }\bibfield
  {title} {\bibinfo {title} {The extraction of work from quantum coherence},\
  }\href {https://doi.org/10.1088/1367-2630/18/2/023045} {\bibfield  {journal}
  {\bibinfo  {journal} {New J. Phys.}\ }\textbf {\bibinfo {volume} {18}},\
  \bibinfo {pages} {023045} (\bibinfo {year} {2016})}\BibitemShut {NoStop}%
\bibitem [{\citenamefont {Perarnau-Llobet}\ \emph {et~al.}(2017)\citenamefont
  {Perarnau-Llobet}, \citenamefont {B\"aumer}, \citenamefont {Hovhannisyan},
  \citenamefont {Huber},\ and\ \citenamefont {Acin}}]{nogo}%
  \BibitemOpen
  \bibfield  {author} {\bibinfo {author} {\bibfnamefont {M.}~\bibnamefont
  {Perarnau-Llobet}}, \bibinfo {author} {\bibfnamefont {E.}~\bibnamefont
  {B\"aumer}}, \bibinfo {author} {\bibfnamefont {K.~V.}\ \bibnamefont
  {Hovhannisyan}}, \bibinfo {author} {\bibfnamefont {M.}~\bibnamefont
  {Huber}},\ and\ \bibinfo {author} {\bibfnamefont {A.}~\bibnamefont {Acin}},\
  }\bibfield  {title} {\bibinfo {title} {No-go theorem for the characterization
  of work fluctuations in coherent quantum systems},\ }\href
  {https://doi.org/10.1103/PhysRevLett.118.070601} {\bibfield  {journal}
  {\bibinfo  {journal} {Phys. Rev. Lett.}\ }\textbf {\bibinfo {volume} {118}},\
  \bibinfo {pages} {070601} (\bibinfo {year} {2017})}\BibitemShut {NoStop}%
\bibitem [{\citenamefont {Lostaglio}(2018)}]{Contextuality}%
  \BibitemOpen
  \bibfield  {author} {\bibinfo {author} {\bibfnamefont {M.}~\bibnamefont
  {Lostaglio}},\ }\bibfield  {title} {\bibinfo {title} {Quantum fluctuation
  theorems, contextuality, and work quasiprobabilities},\ }\href
  {https://doi.org/10.1103/PhysRevLett.120.040602} {\bibfield  {journal}
  {\bibinfo  {journal} {Phys. Rev. Lett.}\ }\textbf {\bibinfo {volume} {120}},\
  \bibinfo {pages} {040602} (\bibinfo {year} {2018})}\BibitemShut {NoStop}%
\bibitem [{\citenamefont {Wigner}(1932)}]{Wigner}%
  \BibitemOpen
  \bibfield  {author} {\bibinfo {author} {\bibfnamefont {E.}~\bibnamefont
  {Wigner}},\ }\bibfield  {title} {\bibinfo {title} {On the quantum correction
  for thermodynamic equilibrium},\ }\href
  {https://doi.org/10.1103/PhysRev.40.749} {\bibfield  {journal} {\bibinfo
  {journal} {Phys. Rev.}\ }\textbf {\bibinfo {volume} {40}},\ \bibinfo {pages}
  {749} (\bibinfo {year} {1932})}\BibitemShut {NoStop}%
\bibitem [{\citenamefont {Kirkwood}(1933)}]{o_K}%
  \BibitemOpen
  \bibfield  {author} {\bibinfo {author} {\bibfnamefont {J.~G.}\ \bibnamefont
  {Kirkwood}},\ }\bibfield  {title} {\bibinfo {title} {Quantum statistics of
  almost classical assemblies},\ }\href {https://doi.org/10.1103/PhysRev.44.31}
  {\bibfield  {journal} {\bibinfo  {journal} {Phys. Rev.}\ }\textbf {\bibinfo
  {volume} {44}},\ \bibinfo {pages} {31} (\bibinfo {year} {1933})}\BibitemShut
  {NoStop}%
\bibitem [{\citenamefont {Dirac}(1945)}]{o_D}%
  \BibitemOpen
  \bibfield  {author} {\bibinfo {author} {\bibfnamefont {P.~A.~M.}\
  \bibnamefont {Dirac}},\ }\bibfield  {title} {\bibinfo {title} {On the analogy
  between classical and quantum mechanics},\ }\href
  {https://doi.org/10.1103/RevModPhys.17.195} {\bibfield  {journal} {\bibinfo
  {journal} {Rev. Mod. Phys.}\ }\textbf {\bibinfo {volume} {17}},\ \bibinfo
  {pages} {195} (\bibinfo {year} {1945})}\BibitemShut {NoStop}%
\bibitem [{\citenamefont {Margenau}\ and\ \citenamefont {Hill}(1961)}]{o_MH}%
  \BibitemOpen
  \bibfield  {author} {\bibinfo {author} {\bibfnamefont {H.}~\bibnamefont
  {Margenau}}\ and\ \bibinfo {author} {\bibfnamefont {R.~N.}\ \bibnamefont
  {Hill}},\ }\bibfield  {title} {\bibinfo {title} {Correlation between
  measurements in quantum theory},\ }\href {https://doi.org/10.1143/PTP.26.722}
  {\bibfield  {journal} {\bibinfo  {journal} {Prog. Theor. Phys.}\ }\textbf
  {\bibinfo {volume} {26}},\ \bibinfo {pages} {722} (\bibinfo {year} {1961})},\
  \Eprint
  {https://arxiv.org/abs/https://academic.oup.com/ptp/article-pdf/26/5/722/5454875/26-5-722.pdf}
  {https://academic.oup.com/ptp/article-pdf/26/5/722/5454875/26-5-722.pdf}
  \BibitemShut {NoStop}%
\bibitem [{\citenamefont {Francica}\ and\ \citenamefont
  {Dell'Anna}(2023)}]{QW_Ising}%
  \BibitemOpen
  \bibfield  {author} {\bibinfo {author} {\bibfnamefont {G.}~\bibnamefont
  {Francica}}\ and\ \bibinfo {author} {\bibfnamefont {L.}~\bibnamefont
  {Dell'Anna}},\ }\href@noop {} {\bibinfo {title} {Quasiprobability
  distribution of work in the ising model}} (\bibinfo {year} {2023}),\ \Eprint
  {https://arxiv.org/abs/2302.11255} {arXiv:2302.11255 [quant-ph]} \BibitemShut
  {NoStop}%
\bibitem [{\citenamefont {Santini}\ \emph {et~al.}(2023)\citenamefont
  {Santini}, \citenamefont {Solfanelli}, \citenamefont {Gherardini},\ and\
  \citenamefont {Collura}}]{KDQ_fermionic}%
  \BibitemOpen
  \bibfield  {author} {\bibinfo {author} {\bibfnamefont {A.}~\bibnamefont
  {Santini}}, \bibinfo {author} {\bibfnamefont {A.}~\bibnamefont {Solfanelli}},
  \bibinfo {author} {\bibfnamefont {S.}~\bibnamefont {Gherardini}},\ and\
  \bibinfo {author} {\bibfnamefont {M.}~\bibnamefont {Collura}},\ }\href@noop
  {} {\bibinfo {title} {Work statistics, quantum signatures and enhanced work
  extraction in quadratic fermionic models}} (\bibinfo {year} {2023}),\ \Eprint
  {https://arxiv.org/abs/2302.13759} {arXiv:2302.13759 [quant-ph]} \BibitemShut
  {NoStop}%
\bibitem [{\citenamefont {Lostaglio}\ \emph {et~al.}(2022)\citenamefont
  {Lostaglio}, \citenamefont {Belenchia}, \citenamefont {Levy}, \citenamefont
  {Hernández-Gómez}, \citenamefont {Fabbri},\ and\ \citenamefont
  {Gherardini}}]{w_KD}%
  \BibitemOpen
  \bibfield  {author} {\bibinfo {author} {\bibfnamefont {M.}~\bibnamefont
  {Lostaglio}}, \bibinfo {author} {\bibfnamefont {A.}~\bibnamefont
  {Belenchia}}, \bibinfo {author} {\bibfnamefont {A.}~\bibnamefont {Levy}},
  \bibinfo {author} {\bibfnamefont {S.}~\bibnamefont {Hernández-Gómez}},
  \bibinfo {author} {\bibfnamefont {N.}~\bibnamefont {Fabbri}},\ and\ \bibinfo
  {author} {\bibfnamefont {S.}~\bibnamefont {Gherardini}},\ }\href@noop {}
  {\bibinfo {title} {Kirkwood-dirac quasiprobability approach to quantum
  fluctuations: Theoretical and experimental perspectives}} (\bibinfo {year}
  {2022}),\ \Eprint {https://arxiv.org/abs/2206.11783} {arXiv:2206.11783
  [quant-ph]} \BibitemShut {NoStop}%
\bibitem [{\citenamefont {Ortega}\ \emph {et~al.}(2019)\citenamefont {Ortega},
  \citenamefont {McKay}, \citenamefont {Alhambra},\ and\ \citenamefont
  {Mart\'{\i}n-Mart\'{\i}nez}}]{w_SE}%
  \BibitemOpen
  \bibfield  {author} {\bibinfo {author} {\bibfnamefont {A.}~\bibnamefont
  {Ortega}}, \bibinfo {author} {\bibfnamefont {E.}~\bibnamefont {McKay}},
  \bibinfo {author} {\bibfnamefont {A.~M.}\ \bibnamefont {Alhambra}},\ and\
  \bibinfo {author} {\bibfnamefont {E.}~\bibnamefont
  {Mart\'{\i}n-Mart\'{\i}nez}},\ }\bibfield  {title} {\bibinfo {title} {Work
  distributions on quantum fields},\ }\href
  {https://doi.org/10.1103/PhysRevLett.122.240604} {\bibfield  {journal}
  {\bibinfo  {journal} {Phys. Rev. Lett.}\ }\textbf {\bibinfo {volume} {122}},\
  \bibinfo {pages} {240604} (\bibinfo {year} {2019})}\BibitemShut {NoStop}%
\bibitem [{\citenamefont {Teixid\'o-Bonfill}\ \emph {et~al.}(2020)\citenamefont
  {Teixid\'o-Bonfill}, \citenamefont {Ortega},\ and\ \citenamefont
  {Mart\'{\i}n-Mart\'{\i}nez}}]{w_SE2}%
  \BibitemOpen
  \bibfield  {author} {\bibinfo {author} {\bibfnamefont {A.}~\bibnamefont
  {Teixid\'o-Bonfill}}, \bibinfo {author} {\bibfnamefont {A.}~\bibnamefont
  {Ortega}},\ and\ \bibinfo {author} {\bibfnamefont {E.}~\bibnamefont
  {Mart\'{\i}n-Mart\'{\i}nez}},\ }\bibfield  {title} {\bibinfo {title} {First
  law of quantum field thermodynamics},\ }\href
  {https://doi.org/10.1103/PhysRevA.102.052219} {\bibfield  {journal} {\bibinfo
   {journal} {Phys. Rev. A}\ }\textbf {\bibinfo {volume} {102}},\ \bibinfo
  {pages} {052219} (\bibinfo {year} {2020})}\BibitemShut {NoStop}%
\bibitem [{\citenamefont {Allahverdyan}(2014)}]{w_MH}%
  \BibitemOpen
  \bibfield  {author} {\bibinfo {author} {\bibfnamefont {A.~E.}\ \bibnamefont
  {Allahverdyan}},\ }\bibfield  {title} {\bibinfo {title} {Nonequilibrium
  quantum fluctuations of work},\ }\href
  {https://doi.org/10.1103/PhysRevE.90.032137} {\bibfield  {journal} {\bibinfo
  {journal} {Phys. Rev. E}\ }\textbf {\bibinfo {volume} {90}},\ \bibinfo
  {pages} {032137} (\bibinfo {year} {2014})}\BibitemShut {NoStop}%
\bibitem [{\citenamefont {Solinas}\ and\ \citenamefont
  {Gasparinetti}(2015)}]{w_FCS}%
  \BibitemOpen
  \bibfield  {author} {\bibinfo {author} {\bibfnamefont {P.}~\bibnamefont
  {Solinas}}\ and\ \bibinfo {author} {\bibfnamefont {S.}~\bibnamefont
  {Gasparinetti}},\ }\bibfield  {title} {\bibinfo {title} {Full distribution of
  work done on a quantum system for arbitrary initial states},\ }\href
  {https://doi.org/10.1103/PhysRevE.92.042150} {\bibfield  {journal} {\bibinfo
  {journal} {Phys. Rev. E}\ }\textbf {\bibinfo {volume} {92}},\ \bibinfo
  {pages} {042150} (\bibinfo {year} {2015})}\BibitemShut {NoStop}%
\bibitem [{\citenamefont {Nazarov}\ and\ \citenamefont
  {Kindermann}(2003)}]{o_FCS}%
  \BibitemOpen
  \bibfield  {author} {\bibinfo {author} {\bibfnamefont {Y.~V.}\ \bibnamefont
  {Nazarov}}\ and\ \bibinfo {author} {\bibfnamefont {M.}~\bibnamefont
  {Kindermann}},\ }\bibfield  {title} {\bibinfo {title} {Full counting
  statistics of a general quantum mechanical variable},\ }\href
  {https://doi.org/10.1140/epjb/e2003-00293-1} {\bibfield  {journal} {\bibinfo
  {journal} {The European Physical Journal B - Condensed Matter and Complex
  Systems}\ }\textbf {\bibinfo {volume} {35}},\ \bibinfo {pages} {413}
  (\bibinfo {year} {2003})}\BibitemShut {NoStop}%
\bibitem [{\citenamefont {Francica}(2022{\natexlab{a}})}]{family}%
  \BibitemOpen
  \bibfield  {author} {\bibinfo {author} {\bibfnamefont {G.}~\bibnamefont
  {Francica}},\ }\bibfield  {title} {\bibinfo {title} {Class of
  quasiprobability distributions of work with initial quantum coherence},\
  }\href {https://doi.org/10.1103/PhysRevE.105.014101} {\bibfield  {journal}
  {\bibinfo  {journal} {Phys. Rev. E}\ }\textbf {\bibinfo {volume} {105}},\
  \bibinfo {pages} {014101} (\bibinfo {year} {2022}{\natexlab{a}})}\BibitemShut
  {NoStop}%
\bibitem [{\citenamefont {Francica}(2022{\natexlab{b}})}]{most_general}%
  \BibitemOpen
  \bibfield  {author} {\bibinfo {author} {\bibfnamefont {G.}~\bibnamefont
  {Francica}},\ }\bibfield  {title} {\bibinfo {title} {Most general class of
  quasiprobability distributions of work},\ }\href
  {https://doi.org/10.1103/PhysRevE.106.054129} {\bibfield  {journal} {\bibinfo
   {journal} {Phys. Rev. E}\ }\textbf {\bibinfo {volume} {106}},\ \bibinfo
  {pages} {054129} (\bibinfo {year} {2022}{\natexlab{b}})}\BibitemShut
  {NoStop}%
\bibitem [{\citenamefont {Miller}\ and\ \citenamefont
  {Anders}(2017)}]{Miller_2017}%
  \BibitemOpen
  \bibfield  {author} {\bibinfo {author} {\bibfnamefont {H.~J.~D.}\
  \bibnamefont {Miller}}\ and\ \bibinfo {author} {\bibfnamefont
  {J.}~\bibnamefont {Anders}},\ }\bibfield  {title} {\bibinfo {title}
  {Time-reversal symmetric work distributions for closed quantum dynamics in
  the histories framework},\ }\href {https://doi.org/10.1088/1367-2630/aa703f}
  {\bibfield  {journal} {\bibinfo  {journal} {New J. Phys.}\ }\textbf {\bibinfo
  {volume} {19}},\ \bibinfo {pages} {062001} (\bibinfo {year}
  {2017})}\BibitemShut {NoStop}%
\bibitem [{\citenamefont {Sampaio}\ \emph {et~al.}(2018)\citenamefont
  {Sampaio}, \citenamefont {Suomela}, \citenamefont {Ala-Nissila},
  \citenamefont {Anders},\ and\ \citenamefont {Philbin}}]{w_Bohm}%
  \BibitemOpen
  \bibfield  {author} {\bibinfo {author} {\bibfnamefont {R.}~\bibnamefont
  {Sampaio}}, \bibinfo {author} {\bibfnamefont {S.}~\bibnamefont {Suomela}},
  \bibinfo {author} {\bibfnamefont {T.}~\bibnamefont {Ala-Nissila}}, \bibinfo
  {author} {\bibfnamefont {J.}~\bibnamefont {Anders}},\ and\ \bibinfo {author}
  {\bibfnamefont {T.~G.}\ \bibnamefont {Philbin}},\ }\bibfield  {title}
  {\bibinfo {title} {Quantum work in the bohmian framework},\ }\href
  {https://doi.org/10.1103/PhysRevA.97.012131} {\bibfield  {journal} {\bibinfo
  {journal} {Phys. Rev. A}\ }\textbf {\bibinfo {volume} {97}},\ \bibinfo
  {pages} {012131} (\bibinfo {year} {2018})}\BibitemShut {NoStop}%
\bibitem [{\citenamefont {Sagawa}(2013)}]{sagawa2013}%
  \BibitemOpen
  \bibfield  {author} {\bibinfo {author} {\bibfnamefont {T.}~\bibnamefont
  {Sagawa}},\ }\bibfield  {title} {\bibinfo {title} {Second law-like
  inequalities with quantum relative entropy: An introduction},\ }in\
  \href@noop {} {\emph {\bibinfo {booktitle} {Lectures on quantum computing,
  thermodynamics and statistical physics}}}\ (\bibinfo  {publisher} {World
  Scientific},\ \bibinfo {year} {2013})\ pp.\ \bibinfo {pages}
  {125--190}\BibitemShut {NoStop}%
\bibitem [{\citenamefont {B{\"a}umer}\ \emph {et~al.}(2018)\citenamefont
  {B{\"a}umer}, \citenamefont {Lostaglio}, \citenamefont {Perarnau-Llobet},\
  and\ \citenamefont {Sampaio}}]{book_QW}%
  \BibitemOpen
  \bibfield  {author} {\bibinfo {author} {\bibfnamefont {E.}~\bibnamefont
  {B{\"a}umer}}, \bibinfo {author} {\bibfnamefont {M.}~\bibnamefont
  {Lostaglio}}, \bibinfo {author} {\bibfnamefont {M.}~\bibnamefont
  {Perarnau-Llobet}},\ and\ \bibinfo {author} {\bibfnamefont {R.}~\bibnamefont
  {Sampaio}},\ }\bibinfo {title} {Fluctuating work in coherent quantum systems:
  Proposals and limitations},\ in\ \href
  {https://doi.org/10.1007/978-3-319-99046-0_11} {\emph {\bibinfo {booktitle}
  {Thermodynamics in the Quantum Regime: Fundamental Aspects and New
  Directions}}},\ \bibinfo {editor} {edited by\ \bibinfo {editor}
  {\bibfnamefont {F.}~\bibnamefont {Binder}}, \bibinfo {editor} {\bibfnamefont
  {L.~A.}\ \bibnamefont {Correa}}, \bibinfo {editor} {\bibfnamefont
  {C.}~\bibnamefont {Gogolin}}, \bibinfo {editor} {\bibfnamefont
  {J.}~\bibnamefont {Anders}},\ and\ \bibinfo {editor} {\bibfnamefont
  {G.}~\bibnamefont {Adesso}}}\ (\bibinfo  {publisher} {Springer International
  Publishing},\ \bibinfo {address} {Cham},\ \bibinfo {year} {2018})\ pp.\
  \bibinfo {pages} {275--300}\BibitemShut {NoStop}%
\bibitem [{\citenamefont {Pan}\ \emph {et~al.}(2019)\citenamefont {Pan},
  \citenamefont {Fei}, \citenamefont {Qiu}, \citenamefont {Zhang},\ and\
  \citenamefont {Quan}}]{Pan2019}%
  \BibitemOpen
  \bibfield  {author} {\bibinfo {author} {\bibfnamefont {R.}~\bibnamefont
  {Pan}}, \bibinfo {author} {\bibfnamefont {Z.}~\bibnamefont {Fei}}, \bibinfo
  {author} {\bibfnamefont {T.}~\bibnamefont {Qiu}}, \bibinfo {author}
  {\bibfnamefont {J.-N.}\ \bibnamefont {Zhang}},\ and\ \bibinfo {author}
  {\bibfnamefont {H.~T.}\ \bibnamefont {Quan}},\ }\href@noop {} {\bibinfo
  {title} {Quantum-classical correspondence of work distributions for initial
  states with quantum coherence}} (\bibinfo {year} {2019}),\ \Eprint
  {https://arxiv.org/abs/1904.05378} {arXiv:1904.05378 [quant-ph]} \BibitemShut
  {NoStop}%
\bibitem [{\citenamefont {Beau}\ \emph {et~al.}(2016)\citenamefont {Beau},
  \citenamefont {Jaramillo},\ and\ \citenamefont {Del~Campo}}]{e18050168}%
  \BibitemOpen
  \bibfield  {author} {\bibinfo {author} {\bibfnamefont {M.}~\bibnamefont
  {Beau}}, \bibinfo {author} {\bibfnamefont {J.}~\bibnamefont {Jaramillo}},\
  and\ \bibinfo {author} {\bibfnamefont {A.}~\bibnamefont {Del~Campo}},\
  }\bibfield  {title} {\bibinfo {title} {Scaling-up quantum heat engines
  efficiently via shortcuts to adiabaticity},\ }\bibfield  {journal} {\bibinfo
  {journal} {Entropy}\ }\textbf {\bibinfo {volume} {18}},\ \href
  {https://doi.org/10.3390/e18050168} {10.3390/e18050168} (\bibinfo {year}
  {2016})\BibitemShut {NoStop}%
\bibitem [{\citenamefont {Lee}\ \emph {et~al.}(2021)\citenamefont {Lee},
  \citenamefont {Ha},\ and\ \citenamefont {Jeong}}]{PhysRevE.103.022136}%
  \BibitemOpen
  \bibfield  {author} {\bibinfo {author} {\bibfnamefont {S.}~\bibnamefont
  {Lee}}, \bibinfo {author} {\bibfnamefont {M.}~\bibnamefont {Ha}},\ and\
  \bibinfo {author} {\bibfnamefont {H.}~\bibnamefont {Jeong}},\ }\bibfield
  {title} {\bibinfo {title} {Quantumness and thermodynamic uncertainty relation
  of the finite-time otto cycle},\ }\href
  {https://doi.org/10.1103/PhysRevE.103.022136} {\bibfield  {journal} {\bibinfo
   {journal} {Phys. Rev. E}\ }\textbf {\bibinfo {volume} {103}},\ \bibinfo
  {pages} {022136} (\bibinfo {year} {2021})}\BibitemShut {NoStop}%
\bibitem [{\citenamefont {Fei}\ \emph {et~al.}(2022)\citenamefont {Fei},
  \citenamefont {Chen},\ and\ \citenamefont {Ma}}]{PhysRevA.105.022609}%
  \BibitemOpen
  \bibfield  {author} {\bibinfo {author} {\bibfnamefont {Z.}~\bibnamefont
  {Fei}}, \bibinfo {author} {\bibfnamefont {J.-F.}\ \bibnamefont {Chen}},\ and\
  \bibinfo {author} {\bibfnamefont {Y.-H.}\ \bibnamefont {Ma}},\ }\bibfield
  {title} {\bibinfo {title} {Efficiency statistics of a quantum otto cycle},\
  }\href {https://doi.org/10.1103/PhysRevA.105.022609} {\bibfield  {journal}
  {\bibinfo  {journal} {Phys. Rev. A}\ }\textbf {\bibinfo {volume} {105}},\
  \bibinfo {pages} {022609} (\bibinfo {year} {2022})}\BibitemShut {NoStop}%
\bibitem [{\citenamefont {Fei}\ and\ \citenamefont
  {Quan}(2019)}]{PhysRevResearch.1.033175}%
  \BibitemOpen
  \bibfield  {author} {\bibinfo {author} {\bibfnamefont {Z.}~\bibnamefont
  {Fei}}\ and\ \bibinfo {author} {\bibfnamefont {H.~T.}\ \bibnamefont {Quan}},\
  }\bibfield  {title} {\bibinfo {title} {Group-theoretical approach to the
  calculation of quantum work distribution},\ }\href
  {https://doi.org/10.1103/PhysRevResearch.1.033175} {\bibfield  {journal}
  {\bibinfo  {journal} {Phys. Rev. Res.}\ }\textbf {\bibinfo {volume} {1}},\
  \bibinfo {pages} {033175} (\bibinfo {year} {2019})}\BibitemShut {NoStop}%
\bibitem [{\citenamefont {Gong}\ and\ \citenamefont {Quan}(2015)}]{Gong2015}%
  \BibitemOpen
  \bibfield  {author} {\bibinfo {author} {\bibfnamefont {Z.}~\bibnamefont
  {Gong}}\ and\ \bibinfo {author} {\bibfnamefont {H.~T.}\ \bibnamefont
  {Quan}},\ }\bibfield  {title} {\bibinfo {title} {Jarzynski equality, crooks
  fluctuation theorem, and the fluctuation theorems of heat for arbitrary
  initial states},\ }\href {https://doi.org/10.1103/PhysRevE.92.012131}
  {\bibfield  {journal} {\bibinfo  {journal} {Phys. Rev. E}\ }\textbf {\bibinfo
  {volume} {92}},\ \bibinfo {pages} {012131} (\bibinfo {year}
  {2015})}\BibitemShut {NoStop}%
\bibitem [{\citenamefont {Dorner}\ \emph {et~al.}(2013)\citenamefont {Dorner},
  \citenamefont {Clark}, \citenamefont {Heaney}, \citenamefont {Fazio},
  \citenamefont {Goold},\ and\ \citenamefont {Vedral}}]{Ramsey_t}%
  \BibitemOpen
  \bibfield  {author} {\bibinfo {author} {\bibfnamefont {R.}~\bibnamefont
  {Dorner}}, \bibinfo {author} {\bibfnamefont {S.~R.}\ \bibnamefont {Clark}},
  \bibinfo {author} {\bibfnamefont {L.}~\bibnamefont {Heaney}}, \bibinfo
  {author} {\bibfnamefont {R.}~\bibnamefont {Fazio}}, \bibinfo {author}
  {\bibfnamefont {J.}~\bibnamefont {Goold}},\ and\ \bibinfo {author}
  {\bibfnamefont {V.}~\bibnamefont {Vedral}},\ }\bibfield  {title} {\bibinfo
  {title} {Extracting quantum work statistics and fluctuation theorems by
  single-qubit interferometry},\ }\href
  {https://doi.org/10.1103/PhysRevLett.110.230601} {\bibfield  {journal}
  {\bibinfo  {journal} {Phys. Rev. Lett.}\ }\textbf {\bibinfo {volume} {110}},\
  \bibinfo {pages} {230601} (\bibinfo {year} {2013})}\BibitemShut {NoStop}%
\bibitem [{\citenamefont {Mazzola}\ \emph {et~al.}(2013)\citenamefont
  {Mazzola}, \citenamefont {De~Chiara},\ and\ \citenamefont
  {Paternostro}}]{PhysRevLett.110.230602}%
  \BibitemOpen
  \bibfield  {author} {\bibinfo {author} {\bibfnamefont {L.}~\bibnamefont
  {Mazzola}}, \bibinfo {author} {\bibfnamefont {G.}~\bibnamefont {De~Chiara}},\
  and\ \bibinfo {author} {\bibfnamefont {M.}~\bibnamefont {Paternostro}},\
  }\bibfield  {title} {\bibinfo {title} {Measuring the characteristic function
  of the work distribution},\ }\href
  {https://doi.org/10.1103/PhysRevLett.110.230602} {\bibfield  {journal}
  {\bibinfo  {journal} {Phys. Rev. Lett.}\ }\textbf {\bibinfo {volume} {110}},\
  \bibinfo {pages} {230602} (\bibinfo {year} {2013})}\BibitemShut {NoStop}%
\bibitem [{\citenamefont {Manzano}\ \emph {et~al.}(2022)\citenamefont
  {Manzano}, \citenamefont {Parrondo},\ and\ \citenamefont
  {Landi}}]{PRXQuantum.3.010304}%
  \BibitemOpen
  \bibfield  {author} {\bibinfo {author} {\bibfnamefont {G.}~\bibnamefont
  {Manzano}}, \bibinfo {author} {\bibfnamefont {J.~M.}\ \bibnamefont
  {Parrondo}},\ and\ \bibinfo {author} {\bibfnamefont {G.~T.}\ \bibnamefont
  {Landi}},\ }\bibfield  {title} {\bibinfo {title} {Non-abelian quantum
  transport and thermosqueezing effects},\ }\href
  {https://doi.org/10.1103/PRXQuantum.3.010304} {\bibfield  {journal} {\bibinfo
   {journal} {PRX Quantum}\ }\textbf {\bibinfo {volume} {3}},\ \bibinfo {pages}
  {010304} (\bibinfo {year} {2022})}\BibitemShut {NoStop}%
\bibitem [{\citenamefont {Yunger~Halpern}\ \emph {et~al.}(2016)\citenamefont
  {Yunger~Halpern}, \citenamefont {Faist}, \citenamefont {Oppenheim},\ and\
  \citenamefont {Winter}}]{YungerHalpern2016}%
  \BibitemOpen
  \bibfield  {author} {\bibinfo {author} {\bibfnamefont {N.}~\bibnamefont
  {Yunger~Halpern}}, \bibinfo {author} {\bibfnamefont {P.}~\bibnamefont
  {Faist}}, \bibinfo {author} {\bibfnamefont {J.}~\bibnamefont {Oppenheim}},\
  and\ \bibinfo {author} {\bibfnamefont {A.}~\bibnamefont {Winter}},\
  }\bibfield  {title} {\bibinfo {title} {Microcanonical and resource-theoretic
  derivations of the thermal state of a quantum system with noncommuting
  charges},\ }\href {https://doi.org/10.1038/ncomms12051} {\bibfield  {journal}
  {\bibinfo  {journal} {Nature Communications}\ }\textbf {\bibinfo {volume}
  {7}},\ \bibinfo {pages} {12051} (\bibinfo {year} {2016})}\BibitemShut
  {NoStop}%
\bibitem [{\citenamefont {Majidy}\ \emph {et~al.}(2023)\citenamefont {Majidy},
  \citenamefont {Braasch}, \citenamefont {Lasek}, \citenamefont {Upadhyaya},
  \citenamefont {Kalev},\ and\ \citenamefont
  {Halpern}}]{majidy2023noncommuting}%
  \BibitemOpen
  \bibfield  {author} {\bibinfo {author} {\bibfnamefont {S.}~\bibnamefont
  {Majidy}}, \bibinfo {author} {\bibfnamefont {W.~F.}\ \bibnamefont {Braasch}},
  \bibinfo {author} {\bibfnamefont {A.}~\bibnamefont {Lasek}}, \bibinfo
  {author} {\bibfnamefont {T.}~\bibnamefont {Upadhyaya}}, \bibinfo {author}
  {\bibfnamefont {A.}~\bibnamefont {Kalev}},\ and\ \bibinfo {author}
  {\bibfnamefont {N.~Y.}\ \bibnamefont {Halpern}},\ }\href@noop {} {\bibinfo
  {title} {Noncommuting conserved charges in quantum thermodynamics and
  beyond}} (\bibinfo {year} {2023}),\ \Eprint
  {https://arxiv.org/abs/2306.00054} {arXiv:2306.00054 [quant-ph]} \BibitemShut
  {NoStop}%
\bibitem [{\citenamefont {Conway}(2007{\natexlab{a}})}]{Conway_6}%
  \BibitemOpen
  \bibfield  {author} {\bibinfo {author} {\bibfnamefont {J.~B.}\ \bibnamefont
  {Conway}},\ }\bibinfo {title} {Normal operators on hilbert space},\ in\ \href
  {https://doi.org/10.1007/978-1-4757-4383-8_9} {\emph {\bibinfo {booktitle} {A
  Course in Functional Analysis}}}\ (\bibinfo  {publisher} {Springer New
  York},\ \bibinfo {address} {New York, NY},\ \bibinfo {year} {2007})\ pp.\
  \bibinfo {pages} {255--302}\BibitemShut {NoStop}%
\bibitem [{\citenamefont {Conway}(2007{\natexlab{b}})}]{Conway_2}%
  \BibitemOpen
  \bibfield  {author} {\bibinfo {author} {\bibfnamefont {J.~B.}\ \bibnamefont
  {Conway}},\ }\bibinfo {title} {Operators on hilbert space},\ in\ \href
  {https://doi.org/10.1007/978-1-4757-4383-8_2} {\emph {\bibinfo {booktitle} {A
  Course in Functional Analysis}}}\ (\bibinfo  {publisher} {Springer New
  York},\ \bibinfo {address} {New York, NY},\ \bibinfo {year} {2007})\ pp.\
  \bibinfo {pages} {26--62}\BibitemShut {NoStop}%
\end{thebibliography}%

\end{document}